\documentclass[journal=inoraj,manuscript=article,layout=twocolumn]{achemso}

\usepackage[version=3]{mhchem} 
\usepackage{txfonts}
\usepackage{color}
\usepackage{hyperref}

\hypersetup{
    colorlinks=true,
    linkcolor=blue,
    filecolor=magenta,      
    urlcolor=cyan,
    pdftitle={Charge-Transfer Complex k-(BEST)2Cu2(CN)3 Analogous to Organic Spin Liquid Candidate},
    pdfpagemode=FullScreen,
    }

\newcommand{\ET}{$\kappa$-ET-CN}
\newcommand{\BEDSe}{$\kappa$-BEST-CN}

\author{Takuya Kobayashi}
\affiliation[Saitama University]
{Graduate School of Science and Engineering, Saitama University, Saitama, 338-8570, Japan }
\altaffiliation{Research and Development Bureau, Saitama University, Saitama 338-8570, Japan}
\email{tkobayashi@mail.saitama-u.ac.jp}
\author{Kent Andrew Sakurai}
\affiliation[Saitama University]
{Graduate School of Science and Engineering, Saitama University, Saitama, 338-8570, Japan }
\author{Shinji Michimura}
\affiliation[Saitama University]
{Graduate School of Science and Engineering, Saitama University, Saitama, 338-8570, Japan }
\author{Hiromi Taniguchi}
\affiliation[Saitama University]
{Graduate School of Science and Engineering, Saitama University, Saitama, 338-8570, Japan }

\title{Charge-Transfer Complex $\kappa$-(BEST)$_2$Cu$_2$(CN)$_3$ Analogous to Organic Spin Liquid Candidate}

\begin{document}

\begin{abstract}
We report the structural, electrical, and magnetic properties of the organic conductor $\kappa$-(BEST)$_2$Cu$_2$(CN)$_3$ (BEST: bis(ethylenediseleno)-tetrathiafulvalene; abbreviated as \BEDSe), which is isostructural with the quantum spin liquid candidate $\kappa$-(ET)$_2$Cu$_2$(CN)$_3$ (ET: bis-(ethylenedithio)tetrathiafulvalene; abbreviated as \ET).
Resistivity measurements demonstrate that \BEDSe\ exhibits semiconducting behavior, governed by the same conducting mechanism as \ET.
Under a pressure of $\sim 0.1$~GPa, \BEDSe\ undergoes a superconducting transition with an onset temperature of $\sim 4$~K. 
From the comparison of the critical pressures of superconductivity between \ET\ and \BEDSe, \BEDSe\ can be regarded as a chemically pressurized analogue of \ET. 
Therefore, \BEDSe, in which only the effective pressure changes without altering the anion structure, is considered a valuable reference material for elucidating the enigmatic properties observed in \ET. Furthermore, the spin susceptibility of \BEDSe\ is slightly larger than that of \ET\ and shows weaker temperature dependence, which cannot be explained by the localized spin model. This behavior clarifies the anomalous magnetic properties of a system with frustration near the Mott transition, serving to stimulate future theoretical research.
\end{abstract}

\section{Introduction}

\begin{figure}
\begin{center}
\includegraphics[width=\columnwidth]{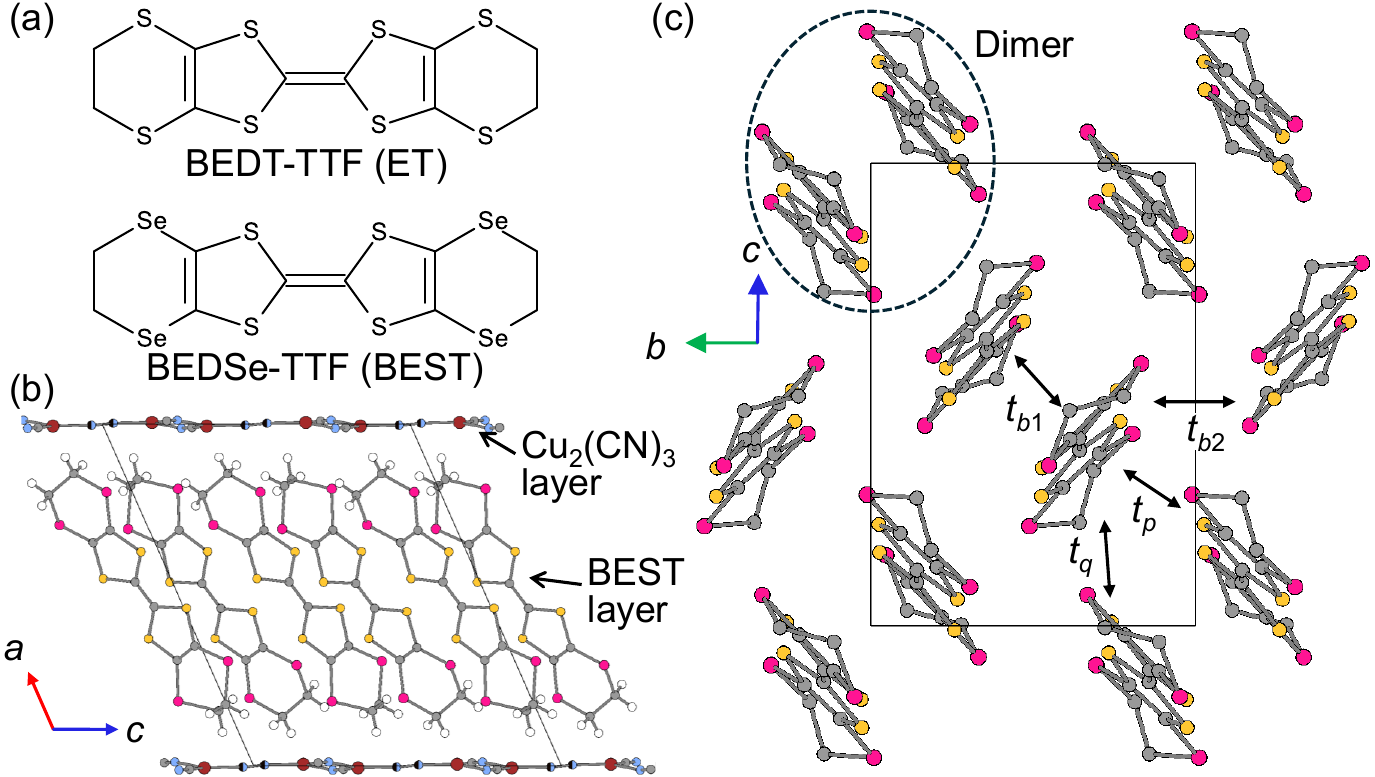}
\end{center}
\caption{
(a) Molecular structures of BEDT-TTF (ET) and BEDSe-TTF (BEST).  
(b) Crystal structure of \BEDSe\ viewed along the $b$ axis. 
(c) $\kappa$-type BEST packing pattern projected along the $a$ axis. 
Calculated transfer integrals (meV) are $t_{b1} = 242.2$, $t_{b2} = 116.9$, $t_{p} = 82.7$, and $t_{q} = -25.3$.
Solid lines and dashed circles represent the unit cell and dimer, respectively.
}
\label{structure}
\end{figure}

The electronic properties induced by the frustration effects are diverse and complex, and have therefore been extensively studied both experimentally and theoretically using various materials and models \cite{Ramirez1994, Balents2010, Savary2017}. 
Quasi-two-dimensional BEDT-TTF-based organic conductors are among the most suitable systems for investigating frustration effects [BEDT-TTF: bis(ethylenedithio)tetrathiafulvalene shown in Fig.~\ref{structure}(a) (abbreviated as ET)] because ET molecules often form a frustrated two-dimensional triangular lattice, and its anisotropy can be tuned through the chemical modifications of the constituents \cite{Shimizu2003, Kagawa2013, Sato2014, Hiramatsu2015}.
$\kappa$-(ET)$_2$Cu$_2$(CN)$_3$ (hereafter \ET) was the first material proposed as a candidate for quantum spin liquid (QSL) \cite{Shimizu2003}, and has attracted considerable attention over the years \cite{Pustogow2022}.
Detailed investigations of \ET\ have revealed several phenomena that cannot be explained by the simple frustrated triangular lattice model. 
These include conflicting results regarding its low-energy gap structure \cite{Yamashita2009, Yamashita2008}, the emergence of a dielectric response in a spin-dominated system \cite{Abdel-Jawad2010}, and a $6$~K anomaly appearing in both the charge and lattice degrees of freedom \cite{Manna2010, Kobayashi2020, Matsuura2022}.
Furthermore, a valence bond solid state has recently been reported \cite{Miksch2021}, indicating that the electronic state of \ET\ remains not fully understood. 

To advance research on QSL and to elucidate the unresolved properties of \ET, alternative QSL candidates have been explored through anion substitution \cite{Hiramatsu2017, Yoshida2019a, Tomeno2020}. 
However, in \ET, the influence of structural disorder caused by positional disorder in the cyano groups on the electronic state has also been discussed \cite{Pinteric2014, Dressel2016}. 
To clarify whether the physical properties observed in \ET\ are intrinsic to the QSL state or originate from the unique structure of the anion, it is desirable to vary the anisotropy of triangular lattice and/or the degree of electron correlation while preserving the structure of the Cu$_2$(CN)$_3$ anion. 

Although anisotropic pressure effects and molecular substitution with BEDT-STF [bis-(ethylenedithio)diselenadithiafulvalene] have been investigated for \ET \cite{Shimizu2003a, Pustogow2021, Saito2021}, tuning the electronic properties finely near the QSL phase remains challenging. 
This is because even small pressures or low substitution ratios tend to induce metallization.
Here, we focused on molecular substitution with BEDSe-TTF [bis(ethylenediseleno)tetrathiafulvalene (abbreviated as BEST)]  (Fig.~\ref{structure}(a)).
Because the outer chalcogen atoms of ET possess small electron densities \cite{Mori1984}, selenium substitutions in such outer sites 
allows finer pressure control compared to BEDT-STF molecular substitution.
In fact, replacing ET in $\kappa$-(ET)$_2$Cu[N(CN)$_2$]Br by BEST results in a negative pressure effect equivalent to less than $0.15$~GPa \cite{Sakata1998,Imajo2022}.
In this paper, we report the synthesis of $\kappa$-(BEST)$_2$Cu$_2$(CN)$_3$ (hereafter \BEDSe) and discuss the effect of donor molecule substitution by comparing its crystal structure, electrical conductivity, and magnetic properties with those of \ET.

\section{Experiments}
The single crystals of \BEDSe\ were synthesized by the electrochemical oxidation of BEST in 1,1,2-trichloroethane and $10$~\% volume of ethanol in the presence of electrolytes. 
Typically, BEST ($29$~mg, $0.050$~mmol) was placed in the anode compartment of an H-shaped electrochemical cell, while a mixture of electrolytes, CuCN ($22$~mg, $0.25$~mmol), KCN ($16$~mg, $0.25$~mmol), and 18-crown-6 ether ($66$~mg, $0.25$~mmol), was added to the cathode compartment. 
Electrocrystallization was carried out at $30$~$^{\circ}$C with a constant current of $0.2$~$\mu$A. 
No alternative polymorphs with different crystal structures were observed under these conditions, indicating that \BEDSe\ is the only crystalline phase obtained by this electrochemical oxidation method.

We performed single-crystal x-ray diffraction analyses on \BEDSe\ at $300$~K and $150$~K using a Bruker D8 QUEST ECO diffractometer equipped with a Bruker PHOTON III detector and a TRIUMPH curved graphite monochromator with Mo $K_\alpha$ radiation ($\lambda$ = $0.71073$~\AA). 
Data reduction and correction were carried out using the APEX5 software suite \cite{bruker2022apex5}. 
The crystal structure was solved using SHELXT program \cite{Sheldrick2015a} and refined by weighted full-matrix least-squares methods on $F^2$ using SHELXL program \cite{Sheldrick2015}, as implemented in Olex2 \cite{Dolomanov2009}.
We observed that many \BEDSe\ crystals exhibited twinning. 
Even the small single crystal selected for structural analysis contained a minor twinning fraction of $\sim 1$\% \cite{Crystallographic}.
The structure was refined with the non-merohedral twin law (1 0 1 / 0 –1 0 / 0 0 –1). Inclusion of the twin model reduced $R_1$ from 4.52\% to 3.71\% ($wR_2$ from 12.80\% to 8.91\%) at $300$~K and from 3.73\% to 3.50\% ($wR_2$ from 9.30\% to 8.31\%) at $150$~K.
Such twinning behavior has also been reported in \ET\ \cite{Komatsu1991, Geiser1991a} and is considered characteristic of charge transfer complexes consisting of Cu$_2$(CN)$_3^{-}$ anions. 

Band structure and intermolecular overlap integrals $S$ were calculated using the extended H\"{u}ckel method and the tight-binding band approximation \cite{Mori1984, Huckelparameter}.
Transfer integrals $t$ were estimated from the relation $t=ES$, where $E$ is the energy level of the highest occupied molecular orbital taken as $-10.0$~eV.

Electrical resistivity measurements were performed using Quantum Design PPMS with the conventional four-probe method.
Gold wires ($25$~$\mu$m in diameter) were attached using carbon paste.
Resistivity under pressures was measured using a Be-Cu/Ni-Cr-Al hybrid piston-cylinder pressure cell, as described in the literature \cite{Uwatoko2002, Taniguchi2010}, with Daphne 7373 oil used as a pressure-transmitting medium \cite{Murata1997}.
For the pressure range of $0.11$--$1.35$~GPa, an inner cylinder with a diameter of $4.5$~mm was employed. 
The applied pressure was determined from the superconducting transition temperature of a lead manometer \cite{Eiling1981}. 
To ensure the reproducibility of the superconducting transition in \BEDSe\ and to allow for more precise pressure control, an inner cylinder with a diameter of $8.0$~mm was used for the pressure range of $0.09$--$0.18$~GPa. 
In this experiment, in-situ pressure determination using the lead manometer was unsuccessful. 
Therefore, the sharp insulator--metal transition observed in the low-pressure region was used to estimate the applied pressure, based on linear interpolation from the relationship between the insulator--metal transition pressure and the pressure values determined using the $4.5$~mm device. 

Magnetic susceptibility measurements were performed on randomly oriented crystals of \BEDSe\ and \ET\ using a superconducting quantum interference device (SQUID) magnetometer (Quantum Design MPMS XL-7) under a magnetic field of $0.5$~T. 
The core diamagnetic contributions of \BEDSe\ and \ET\ are $-5.18 \times 10^{-4}$ and $-4.37 \times 10^{-4}$~emu/mol (value for \ET\ was taken from Ref.~\cite{Shimizu2003}), respectively, and we discuss the spin susceptibility by subtracting them from the measured magnetic susceptibility. 
The sample masses were $10.3$~mg for \BEDSe\ and $14.6$~mg for \ET. 

\section{Results and discussion}

\subsection{Crystal structure}
Table~\ref{t1} lists the lattice parameters of \BEDSe\ for the crystal structure refined in the space group $P2_1/c$ \cite{Spacegroup}  (see Ref.~[\cite{Crystallographic}] for details of the structural refinement). 
For comparison, the parameters of \ET\ are also shown \cite{Jeschke2012}, indicating that \BEDSe\ is isostructural with \ET. 
The asymmetric unit contains one crystallographically independent BEST molecule. 
The only one of its two terminal ethylene groups exhibits orientational disorder with $79$\% adopting a staggered conformation at $300$~K.
We confirmed that they become ordered at $150$~K, which is consistent with that observed in \ET \cite{Jeschke2012} (Table~\ref{t1}). 

The crystal structure of \BEDSe\ is illustrated in Fig.~\ref{structure}(b) and \ref{structure}(c). 
This compound has a layered structure, in which BEST donor layers and Cu$_2$(CN)$_3^-$ anion layers alternate along the $a$ axis. 
Within the BEST layers, the donor molecules are arranged in a $\kappa$-type configuration. 
In the anion layers, disorder is observed in the CN bonds along the $c$ axis, with the occupancy probabilities of C and N assumed to be  $50$\%.
These structural features are the same as those reported for \ET \cite{Geiser1991a, Jeschke2012}, and microscopic measurements have also confirmed equal occupancies of C and N in this salt \cite{Kobayashi2020}.

The unit cell volume of \BEDSe\ is $4.0$\% larger than that of \ET, and each lattice constant of \BEDSe\ is also larger, with no significant anisotropic change (Table~\ref{t1}). 
This might suggest a negative chemical pressure effect. 
On the other hand, the positive pressure effect resulting from the increased overlap of Se orbitals may be more dominant than the increase in volume. 
The actual pressure effect will be discussed based on resistivity measurements.

\begin{table*}
\caption{Lattice parameters of \BEDSe\ and \ET.
The staggered represents the occupancy of the staggered conformation of the ethylene group in the BEST molecule.
}
\label{t1}
\begin{center}
\begin{tabular}{lcccc}
\hline
 & \BEDSe & \BEDSe & \ET & \ET \\
 \hline
 $T$ (K)& 300& 150& 300&150\\
$a$\ (\AA) & 16.4504(6)& 16.4171(11)& 16.0919(3) & 16.0703(3)\\
$b$\ (\AA) & 8.6983(3)& 8.6876(6)& 8.5722(2) & 8.5664(2)\\
$c$\ (\AA) & 13.5492(5)& 13.4152(9)& 13.3889(2) & 13.2698(3)\\
$\beta$\ (deg.)  & 113.9090(10)& 114.769(2)& 113.4060(10) & 114.609(1)\\
$V$\ (\AA$^3$) & 1772.40(11)& 1737.3(2)& 1694.93(6) & 1660.72(6)\\
staggered (\%) & 79 & 100& 77 & 100\\
CCDC number & 2442311 & 2485419 & 850022 & 850025 \\
Reference & This work & This work&  [40] & [40] \\
\hline
\end{tabular}
\end{center}
\label{lattice}
\end{table*}

\subsection{Band structure}
\begin{figure}[t]
\begin{center}
\includegraphics[width=\columnwidth]{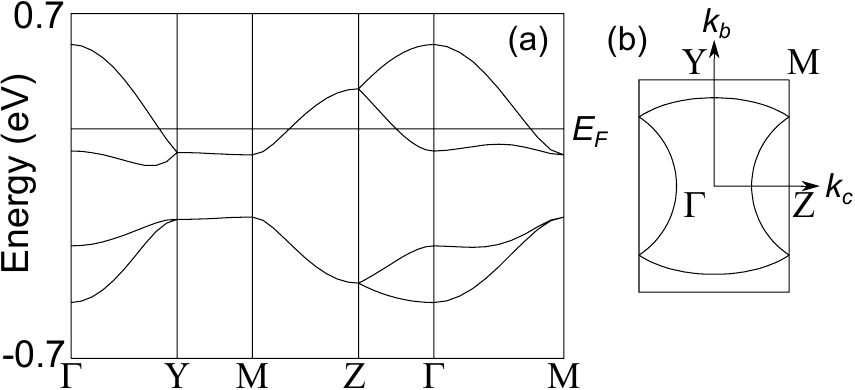}
\end{center}
\caption{(a) Band dispersion and (b) Fermi surface of \BEDSe. 
}
\label{band}
\end{figure}

Using the crystal structure determined at $300$~K, we performed the extended H\"{u}ckel and tight-binding calculations \cite{Mori1984, Huckelparameter}.
From these calculations, we obtained the transfer integrals for \BEDSe: $t_{b1} = 242.2$, $t_{b2} = 116.9$, $t_{p} = 82.7$, and $t_{q} = -25.3$~meV, with the definitions  displayed in Fig.~\ref{structure}(c). 
$t_{\rm b1}$ is larger than the other values, indicating that two molecules form a dimer, which is characteristic of the $\kappa$-type salts. 
Dimerization splits the energy band into upper and lower bands, as shown in Fig.~\ref{band}(a), resulting in an effective half-filled band structure.
The calculation also demonstrates the two-dimensional Fermi surface consisting of a hole-like sheet around the $Z$--point and an electron-like sheet parallel to $M$--$Y$ [Fig.~\ref{band}(b)] as observed in \ET\ \cite{Komatsu1996}.

The obtained transfer integrals and band dispersion provide an estimate of the electron correlation $U/W = 0.91$, where $W$ is the bandwidth estimated from the upper branch of the dispersion, and $U$ is the on-site Coulomb repulsion defined as $U = 2t_{b1}$ \cite{Kanoda1997c}. 
This ratio indicates that \BEDSe\ is located near the Mott boundary.
The anisotropy of the triangular lattice $t'/t$ was estimated to be $1.08$ by the dimer approximation, with $t' = |t_{b2}|/2$ and $t = (|t_p| + |t_q|)/2$. 
The obtained values of $U/W$ and $t'/t$ are close to those of the \ET\ ($U/W \sim 0.90-0.93$ and $t'/t \sim 1.06-1.09$ at room temperature) \cite{Komatsu1996, Shimizu2003, Hiramatsu2015}, suggesting that the two salts are in similar electronic states.

\subsection{Electrical resistivity}
\begin{figure}[t]
\begin{center}
\includegraphics[width=0.8\columnwidth]{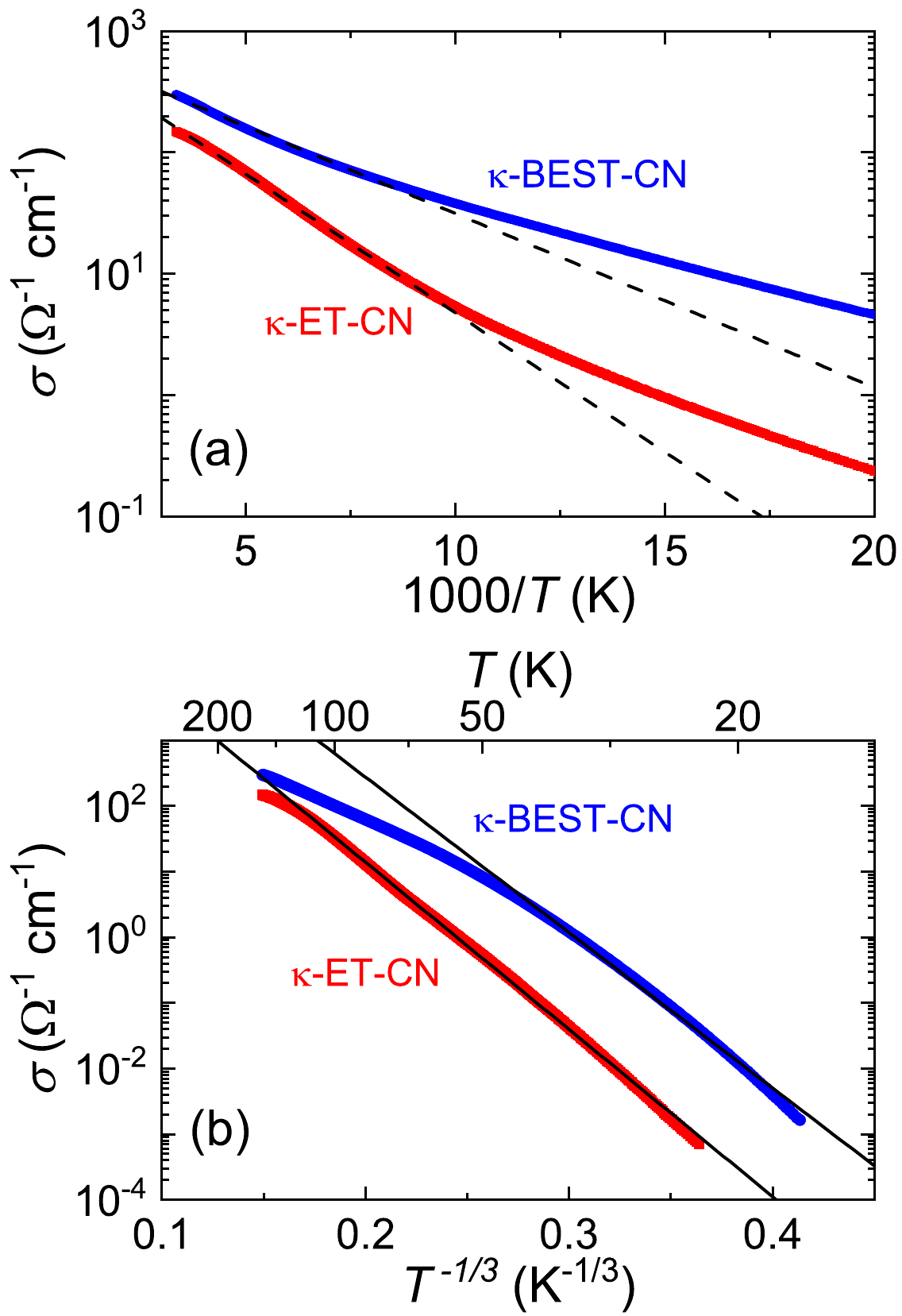}
\end{center}
\caption{
(a) In-plane conductivities $\sigma$ of \BEDSe\ and \ET\ as a function of $T^{-1}$. The dashed lines are the fitting curves with the NNH model for $\sigma$ above $\sim 100$~K. 
(b) $\sigma$ as a function of $T^{-1/3}$, where the solid lines are fitting curves with the two-dimensional VRH model for the low-temperature data. 
}
\label{resistivity}
\end{figure}

The electrical conductivities $\sigma$ of \ET\ and \BEDSe\ are plotted in Fig.~\ref{resistivity}(a) as a function of inverse temperature $1/T$. 
The current was applied parallel to the conduction plane.
The $\sigma$ at room temperature is $3.0\times10^2$~$\Omega^{-1}$ cm$^{-1}$ and $1.5\times10^2$~$\Omega^{-1}$ cm$^{-1}$ for \BEDSe\ and \ET, respectively, which are suggested to be almost the same, considering the error in estimating the sample size.
The $\sigma$ of both salts shows semiconducting behavior over the entire temperature range but does not match a simple thermally activated model. 
The results for \ET\ are consistent with previous reports \cite{Kawamoto2004a, Pinteric2014, Saito2018, Culo2019}, and the change in $\sigma$ with decreasing temperature has been understood by a crossover from nearest-neighbor-hopping (NNH) to two-dimensional variable-range-hopping (VRH). 

As shown by the dashed lines in Fig.~\ref{resistivity}(a), the $\sigma$ of both salts at high temperatures can be described by the NNH model $\sigma \propto \exp(-\Delta/T)$, where $\Delta$ is the activation energy. 
Fits to the temperature dependence of $\sigma$ above approximately $100$~K yield $\Delta = 528$~K for \ET\ and $330$~K for \BEDSe. 
In \BEDSe, there appears to be a deviation from the NNH model at higher temperatures.

In the VRH model in two dimensions, the $\sigma$ is expressed as $\sigma \propto \exp[-(T_0/T)^{1/(d+1)}]$ with $d = 2$, where $k_B T_0$ is the corresponding activation energy and $k_B$ is the Boltzmann constant. 
To analyze the temperature dependence of $\sigma$ using this model, the $\sigma$ of both salts are plotted as a function of $T^{-1/3}$ in Fig.~\ref{resistivity}(b), together with the fitting curves based on the VRH model (solid lines).
The temperature dependence of $\sigma$ at low temperatures is well reproduced by adopting $k_B T_0 = 17.0$~eV for \ET\ and $k_B T_0 = 13.7$~eV for \BEDSe. 

From these analyses, both $\Delta$ and $k_B T_0$ for \BEDSe\ are smaller than those for \ET. 
This suggests that the increased bandwidth in \BEDSe\ leads to lower activation energies, placing \BEDSe\ on the higher-pressure side relative to \ET. 
We see that the VRH model fits the data over a wide temperature range for \ET, whereas for \BEDSe, the model appears valid only below approximately $40$~K. 
This could indicate a broader crossover region from NNH to VRH. 
As discussed for \ET \cite{Byczuk2005,Culo2019}, the increased bandwidth likely results in the more intermediate conducting behavior suggested by the phase diagram near the Mott--Anderson localization state.

\subsection{Electrical resistivity under pressures}
\begin{figure}[t]
\begin{center}
\includegraphics[width=1\columnwidth]{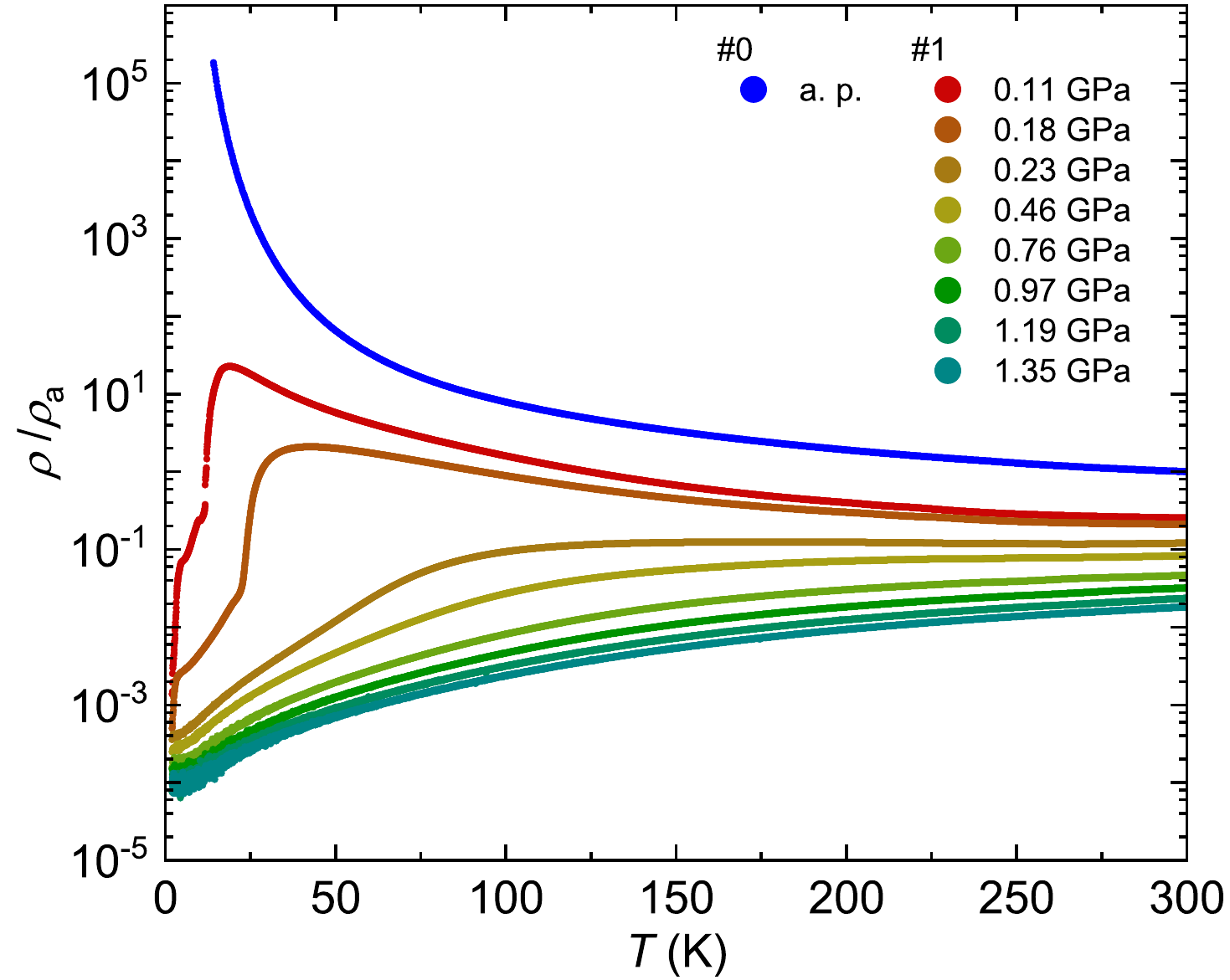}
\end{center}
\caption{
Temperature dependence of out-of-plane resistivity $\rho$ of \BEDSe\ at various pressures normalized by resistivity $\rho_{\rm a}$ under ambient pressure ($0$~GPa) and temperature ($300$~K). 
Samples \#0 and \#1 were measured at ambient pressure (a. p.) and under pressures, respectively.
}
\label{rhoTup}
\end{figure}

Figure~\ref{rhoTup} shows the temperature dependence of the electrical resistivity ($\rho$) of \BEDSe\ at various pressures, measured using a $4.5$~mm-diameter piston-cylinder pressure cell.
The measurements were carried out with the current applied perpendicular to the conduction plane.
The resistivity under ambient and pressurized conditions was measured on different samples (labeled \#0 and \#1, respectively), and all values are normalized by the resistivity at ambient pressure and room temperature ($\rho_{\rm a}$ at $0$~GPa and $300$~K) for comparison. 
As described above, semiconducting behavior was observed at ambient pressure. 
With increasing pressure, the normalized resistivity $\rho/\rho_{\rm a}$ decreases, and at $0.11$~GPa, an insulator--metal transition was observed below $15$~K.
Further application of pressure leads to metallization even at temperatures below $300$~K.

\begin{figure}[t]
\begin{center}
\includegraphics[width=1\columnwidth]{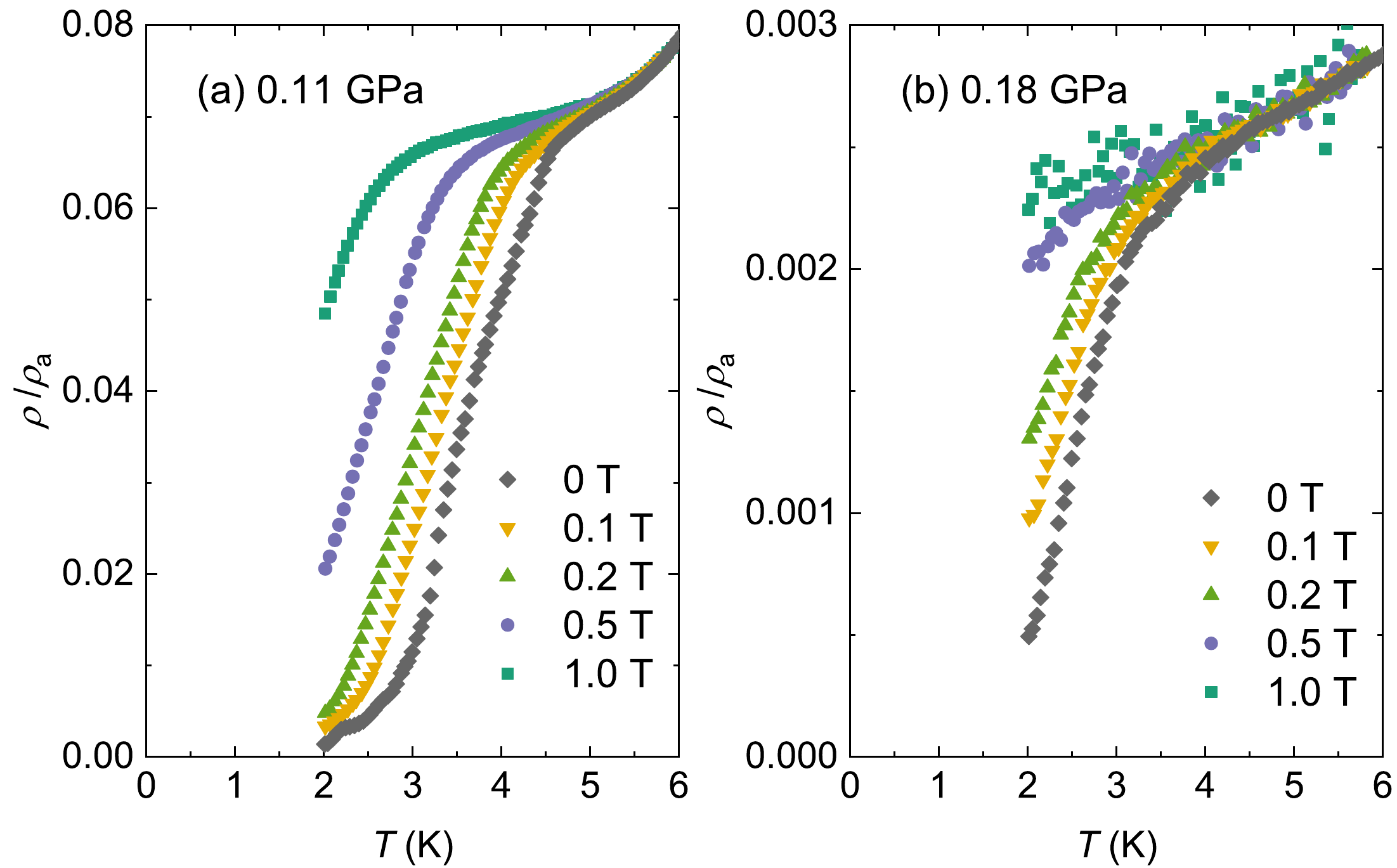}
\end{center}
\caption{
Temperature dependence of normalized resistivity at (a) $0.11$~GPa and (b) $0.18$~GPa under various magnetic fields. 
}
\label{rhoTSC}
\end{figure}

At $0.11$ and $0.18$~GPa, where sharp insulator--metal transitions were observed, drastic decreases in $\rho/\rho_{\rm a}$ were observed below approximately $4.5$~K and $3.5$~K, respectively (Fig.~\ref{rhoTSC}(a), (b)). 
These decreases in $\rho/\rho_{\rm a}$ are suppressed when a magnetic field is applied, indicating the onset of a superconducting transition. 
It is noteworthy that the present observations identify \BEDSe\ as the second example of a BEST-based superconductor, subsequent to $\kappa$-(BEST)$_2$Cu[N(CN)$_2$]Br \cite{Sakata1998}.
The $0.11$~GPa at which \BEDSe\ initiates to exhibit the insulator--metal and superconducting transitions is less than $0.135$--$0.140$~GPa at which \ET\ does \cite{Furukawa2018}. 
This difference signifies that the BEST molecular substitution causes a positive chemical pressure effect.

\begin{figure}[t]
\begin{center}
\includegraphics[width=1\columnwidth]{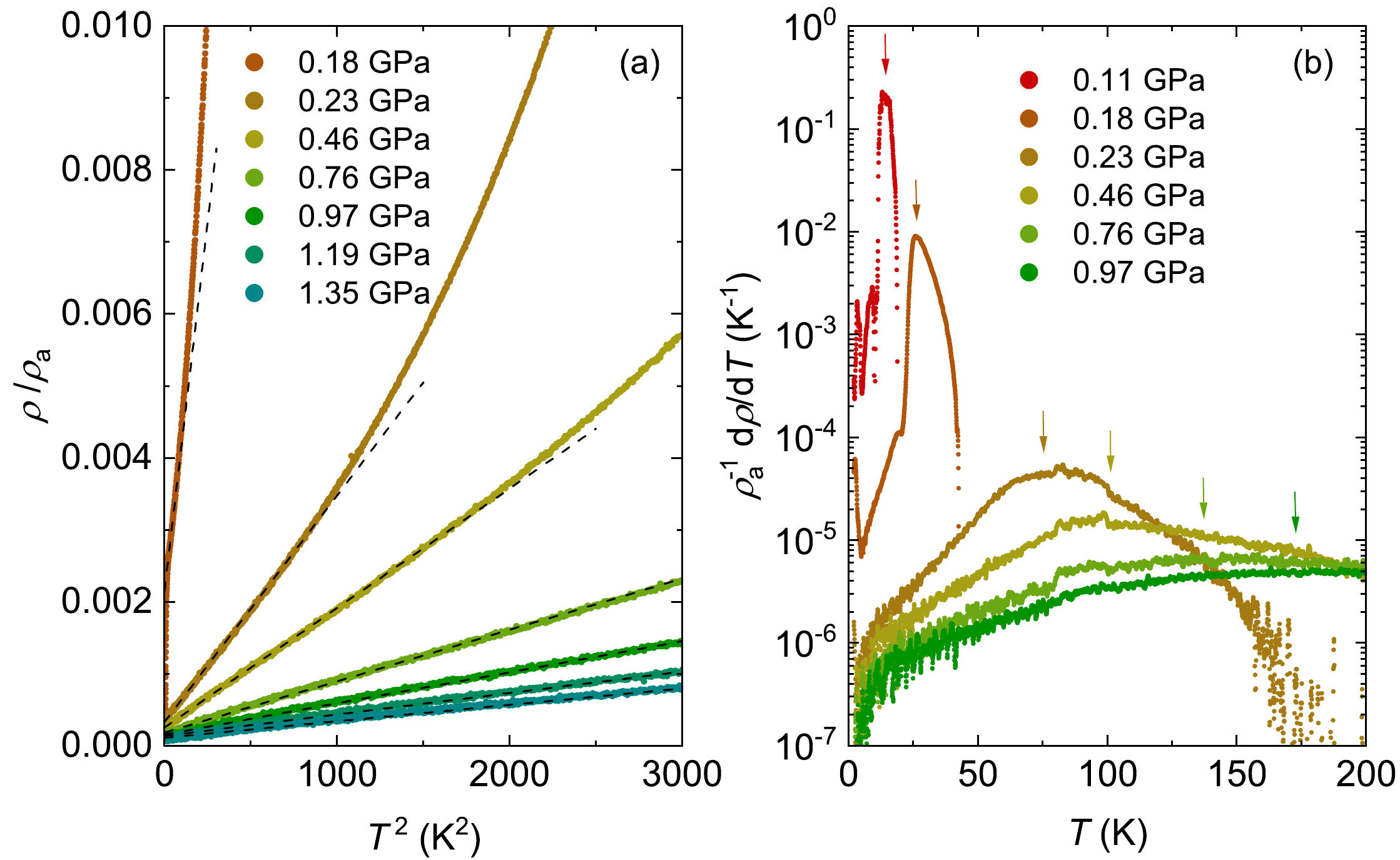}
\end{center}
\caption{
(a) Normalized resistivity as a function of $T^2$ at $0.18$--$1.35$~GPa exhibiting Fermi liquid-like behavior at low temperatures.
(b) Temperature dependence of the first derivative of normalized resistivity at $0.11$--$0.97$~GPa showing peak behavior indicated by arrows. 
}
\label{T2anddRdT}
\end{figure}

Figure~\ref{T2anddRdT}(a) shows the $T^2$ dependence of $\rho/\rho_{\rm a}$ for \BEDSe\ at low temperatures. 
According to the conventional Fermi liquid theory, $\rho$ is expected to be proportional to $T^2$, a relationship already confirmed for \ET \cite{Kurosaki2005}. 
The same behavior is evident in the metallic state of \BEDSe, as indicated by the dashed lines.
Deviations of the data from those lines at lower pressures suggest that the Fermi liquid state is established only at low temperatures. 
The Fermi liquid state is stabilized from higher temperatures as the applied pressure increases. 
In other words, the coherence temperature, $T_{\rm FL}$, defined as the temperature below which $\rho$ follows a $T^2$ dependence, increases with increasing pressure.

The maximum of the temperature derivative of $\rho$ also characterizes the insulator--metal transition, which has been widely discussed in organic conductors \cite{Murata1990, Taniguchi1999, Kurosaki2005, Strack2005, Kobayashi2017}. 
Figure~\ref{T2anddRdT}(b) shows the temperature dependence of $\rho_{\rm a}^{-1}\mathrm{d}\rho/\mathrm{d}T$, with the temperature at which this quantity has a maximum, $T_{\rm IM}$, indicated by the arrows, shifting to higher temperatures as pressure increases. 
Additionally, the maximum of this quantity becomes broader. 
These behaviors are qualitatively similar to those observed in \ET \cite{Kurosaki2005}.

\begin{figure}[t]
\begin{center}
\includegraphics[width=1\columnwidth]{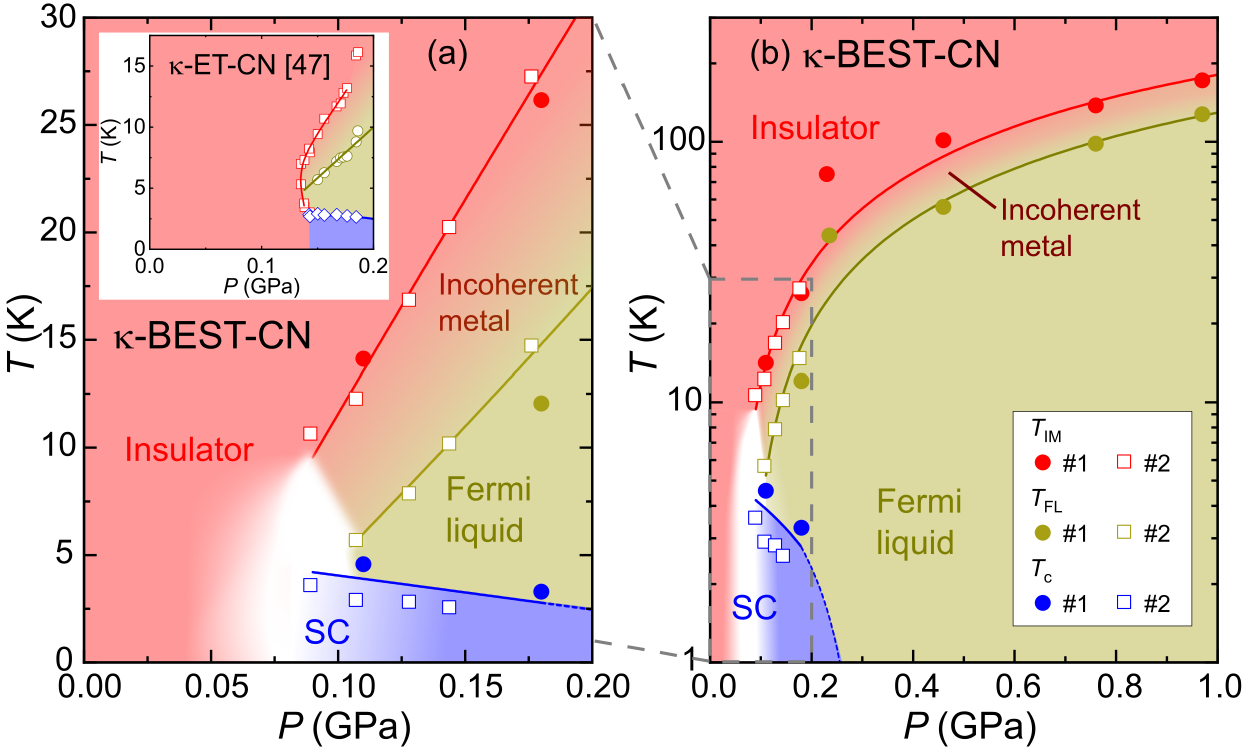}
\end{center}
\caption{
$T$--$P$ phase diagram of \BEDSe\ in (a) a narrow low-pressure region and (b) a high-pressure region.
Closed and open symbols correspond to data measured with samples \#1 and \#2, where currents are applied perpendicular and parallel to the conduction plane, respectively.  
$T_{\rm IM}$, $T_{\rm FL}$, and $T_{\rm c}$ are the insulator--metal transition temperatures determined from the $\mathrm{d}\rho/\mathrm{d}T$ peak, the temperature at which $\rho$ deviates from $T^2$ dependence, and the onset superconducting transition temperature, respectively.
Solid curves are guides to the eye.
Inset shows the $T$--$P$ phase diagram of \ET\ \cite{Furukawa2018}.
}
\label{PT}
\end{figure}

Figure~\ref{PT}(a) and \ref{PT}(b) show the temperature--pressure ($T$--$P$) phase diagram of \BEDSe.
To verify reproducibility, results obtained from a different sample (\#2) using an $8.0$~mm-diameter piston-cylinder pressure cell are also plotted with different symbols, where, in this case, the current was applied parallel to the conduction plane.
In both samples, a superconducting transition was observed following the insulator-metal transition, indicating that the superconducting phase lies below the metallic phase.
For comparison, the phase diagram of \ET\ \cite{Furukawa2018} are shown in the inset of Fig.~\ref{PT}(a). 
All characteristic boundaries in \BEDSe\ are shifted to lower pressure by $\sim 0.05$~GPa, suggesting that \BEDSe\ lies on the higher effective pressure side relative to \ET.

According to the Clausius–Clapeyron relation for a first-order transition, the slope of the phase transition $\mathrm{d}T_{\rm IM}/\mathrm{d}P$ satisfies $\mathrm{d}T_{\rm IM}/\mathrm{d}P =\Delta V/\Delta S$, where $\Delta V$ and $\Delta S$ are the volume and entropy changes at the transition, respectively. 
The positive slope of $\mathrm{d}T_{\rm IM}/\mathrm{d}P$ observed in \BEDSe\ (Fig.~\ref{PT}) indicates that the insulating state has larger entropy than the neighboring metallic state, which is similar to the behavior of \ET\ in the inset of Fig.~\ref{PT}(a).
At low temperatures in \ET, the $\mathrm{d}T_{\rm IM}/\mathrm{d}P$ changes from positive to negative (inset of Fig.~\ref{PT}(a)), suggesting a transition to a state with lower entropy in the insulating phase.
To establish whether \BEDSe\ also exhibits a similar insulator--metal transition at lower temperatures, precise measurements under finely controlled pressures are required and remain a key objective.

\subsection{BEST molecular substitution}
\begin{table*}
\caption{Intermolecular contacts $d$ shorter than the sum of van der Waals radii $R_{ij}$ ($= r_i + r_j$) with $r_\mathrm{C} = 1.7$~\AA, $r_\mathrm{S} = 1.8$~\AA, and $r_\mathrm{Se} = 1.9$~\AA \cite{Batsanov2001}. 
$\delta$ represents $\delta = (d - R_{ij})/R_{ij}$.
}
\begin{center}
\begin{tabular}{l|lcc|lcc}
\hline
 & \multicolumn{3}{c|}{\BEDSe\ ($150$~K)} & \multicolumn{3}{|c}{\ET\ ($150$~K)}\\
 \hline
 & Contact & $d$ (\AA) & $\delta$(\%) & Contact& $d$ (\AA)& $\delta$(\%)\\
$b1$ & (i) S1$\cdots$S2 & 3.5924(9)& -0.2 & (i) S1$\cdots$S2 & 3.6500(3) & +1.4\\
     & (ii) C1$\cdots$C1 & 3.356(4)& -1.3 & (ii) C1$\cdots$C1 & 3.392(2) & -0.2\\
$b2$ & (iii) Se2$\cdots$S4 & 3.6580(8)& -1.1 & (iii) S5$\cdots$S3 & 3.6221(4) & +0.6\\
     & (iv) Se2$\cdots$Se4 & 3.6620(5)&  -3.6 & (iv) S5$\cdots$S7 & 3.6981(5)& +2.7\\
$p$ & (v) C4$\cdots$Se3 & 3.360(3)& -6.7 & (v) C3$\cdots$S8 & 3.407(1)& -2.7\\
   & (vi) C3$\cdots$Se3 & 3.573(3)&  -0.8 & (vi) C4$\cdots$S8& 3.627(1)& +3.6\\
$q$ & (vii) S1$\cdots$Se2 & 3.5726(8)& -3.4 & (vii) S2$\cdots$S5 & 3.5362(5)& -1.8\\
\hline
\end{tabular}
\end{center}
\label{vdW}
\end{table*}
\begin{figure}[t]
\begin{center}
\includegraphics[width=1\columnwidth]{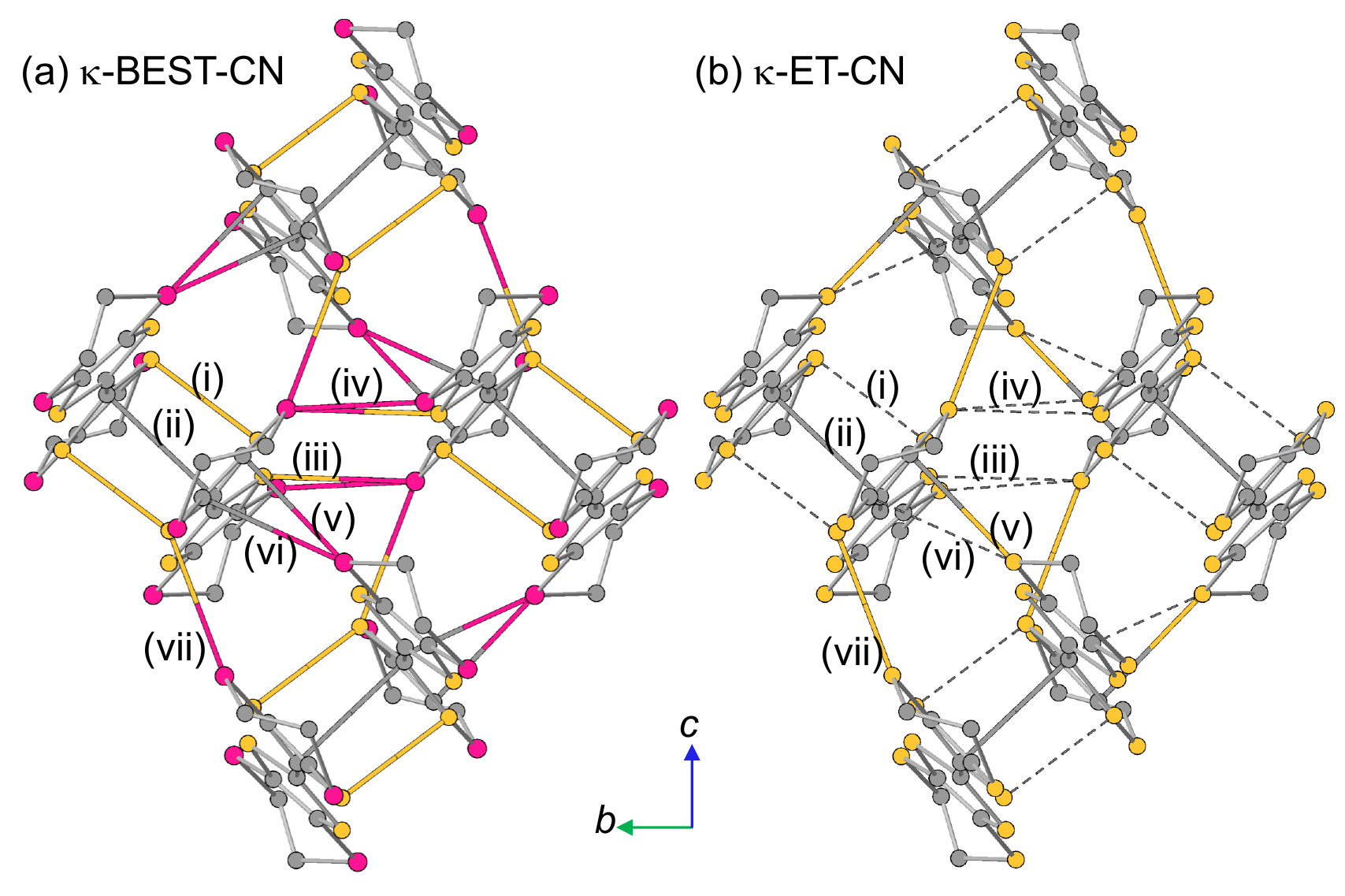}
\end{center}
\caption{
Donor molecular arrangement viewed along the $a$ axis at $150$~K for \BEDSe\ (CCDC No. 2475419) and \ET\ (CCDC No. 850025 \cite{Jeschke2012}).
In \BEDSe, intermolecular contacts shorter than the sum of the van der Waals radii ($R_{ij}$) are shown as thick lines colored according to the atoms involved.
For \ET, the corresponding contacts are indicated by thick lines ($< R_{ij}$) and dashed lines ($> R_{ij}$).
}
\label{contact}
\end{figure}
Selenium substitution in ET molecules causes lattice expansion and increased intermolecular orbital overlap. The former results in a negative chemical pressure effect, while the latter results in a positive one.
For example, BETS (bis(ethylenedithio)tetraselenafulvalene) molecule, in which the inner sulfur atoms of ET are substituted with selenium, affords many metallic and superconducting salts \cite{Kobayashi2004}. On the other hand, substituting the outer sulfur atoms with selenium causes the steric effect \cite{Courcet1998}. 
In fact, many known BEST salts are semiconductors at low temperatures \cite{Wang1989,Lyubovskaya1997,Sakata1998,Clemente-Leon2002}.
Nevertheless, resistivity measurements reveal a positive pressure effect in \BEDSe.
To investigate the origin of this behavior, we analyzed the short contacts within the donor layers of \ET\ and \BEDSe. 
Figure~\ref{contact}(a) shows the intermolecular contacts (thick lines) for \BEDSe\ at $150$~K shorter than the sum of the van der Waals radii, $R_{ij} =  r_{i} + r_{j}$, with the van der Waals radius $r_{i}$($r_{j}$) of atom $i$($j$). 
For comparison, the corresponding contacts for \ET\ at the same temperature are shown in Fig.~\ref{contact}(b) with thick lines ($<R_{ij}$) and dashed lines ($>R_{ij}$) \cite{Jeschke2012}. 
These short contacts $d$ together with the relative distances $\delta = (d-R_{ij})/R_{ij}$, are listed in Table~\ref{vdW} and sorted by the paths of the transfer integrals defined in Fig.~\ref{structure}(c). 
As shown, $\delta$ is smaller for \BEDSe\ than for \ET\ for all contacts.
In particular, $b2$, $p$, and $q$ paths in \BEDSe\ include the short contacts that are more than $3$\% shorter than $R_{ij}$. 
Considering that the universal phase diagram of $\kappa$‑type salts can be understood in terms of $U/W$ \cite{Kanoda1997c}, the slight increase in $U$ ($\propto t_{b1}$) and the significant increase in $W$ ($\propto t_{b2}, t_p, t_q$) explain the observation of the positive pressure effect by BEST substitution.

Many BEST salts are semiconductors, and superconducting $\kappa$‑(ET)$_2$Cu[N(CN)$_2$]Br \cite{Kini1990} and $\beta$‑(ET)$_2$I$_3$ \cite{Yagubskii1984} become semiconducting by BEST substitution \cite{Sakata1998,Wang1989}. 
Therefore, BEST molecules are generally expected to have a negative pressure effect compared to ET molecules.
However, the present resistivity measurements reveal a positive pressure effect. We discuss the origin of this contrasting situation. 
It has been pointed out that in \ET, anion and donor molecules are packed in a key-on-hole relationship \cite{Geiser1991,Yamochi1993, Hiramatsu2015}. 
In \BEDSe, this relationship appears to limit the volume expansion caused by substituting BEST for ET, thereby maintaining relatively short intermolecular distances. 
I$_3^-$ and Cu[N(CN)$_2$]Br$^-$ are discrete and one-dimensional polymer-like anions, respectively. The difference between them and the two-dimensional anion network structure of Cu$_2$(CN)$_3^-$ may qualitatively explain the difference in chemical pressure effects.

\subsection{Spin susceptibility}
\begin{figure}[t]
\begin{center}
\includegraphics[width=0.9\columnwidth]{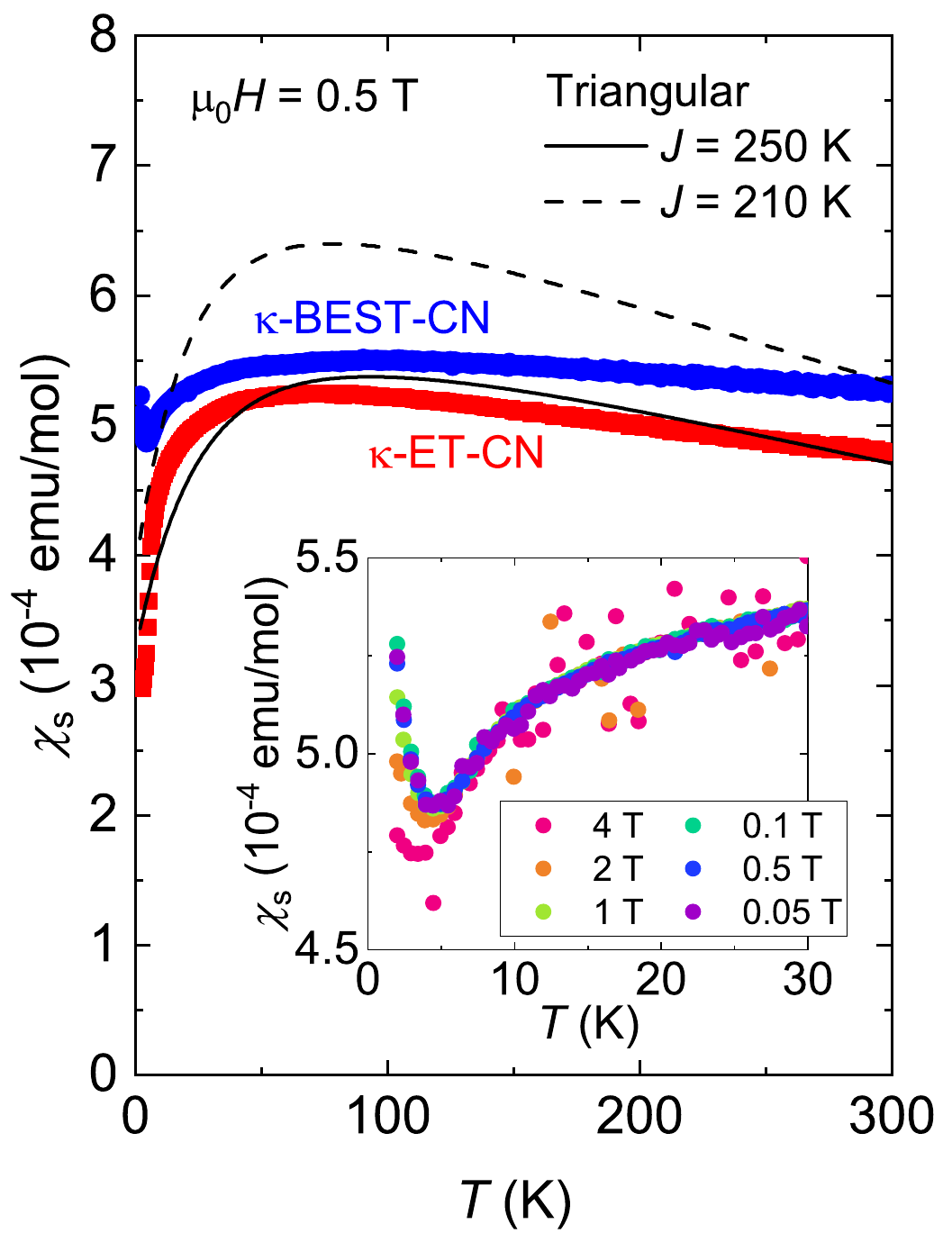}
\end{center}
\caption{
Temperature dependence of the spin susceptibility of polycrystalline \ET\ and \BEDSe\ under the magnetic field of $0.5$~T. 
Solid and dashed curves are the calculated susceptibilities for the spin-$1/2$ Heisenberg antiferromagnetic triangular lattice \cite{Elstner1993} with $J = 250$~K and $J = 210$~K, respectively. 
The inset shows the temperature dependence of the spin susceptibility of \BEDSe\ at various magnetic fields.
}
\label{sus}
\end{figure}

Figure~\ref{sus} shows the temperature dependence of spin susceptibility $\chi_{\rm s}$ for \BEDSe\ and \ET. 
The $\chi_{\rm s}$ of \ET\ increases with decreasing temperature from $300$~K, and has a broad peak at approximately $80$~K, below which it steeply decreases. 
This overall temperature dependence of $\chi_{\rm s}$ is explained by the spin-$1/2$ two-dimensional Heisenberg antiferromagnetic triangular lattice model \cite{Elstner1993} with an exchange interaction of $J$ = $250$~K, as shown by the solid curve. 
These results are consistent with previous studies \cite{Shimizu2003, Saito2018}.
In contrast, $\chi_{\rm s}$ of \BEDSe\ exhibits only weak temperature dependence. 
The room-temperature value of $\chi_{\rm s}$ $= 5.3 \times 10^{-4}$~emu/mol in \BEDSe\ is larger than $\chi_{\rm s} = 4.8 \times 10^{-4}$~emu/mol in \ET. 
In addition, the $\chi_{\rm s}$ value of \BEDSe\ is significantly larger than the typical spin susceptibilities of metallic $\kappa$-type salts  ($\chi_{\rm s} = (4.2-4.5) \times 10^{-4}$~emu/mol) \cite{Kanoda1997c,Kanoda2006}. 
Based on this higher value, we adopted the antiferromagnetic triangular lattice model  with $J$ = $210$~K shown by the dashed curve in the figure, but as can be seen, it deviates significantly from experimental data. 

The positive pressure effect and the almost temperature-independent $\chi_{\rm s}$ are reminiscent of the Pauli paramagnetism expected in the metallic state, but it is inconsistent with the semiconducting behavior of \BEDSe. 
Two models can be considered to explain this. 
The first is that, considering that the $t'/t$ of \BEDSe\ is close to unity, the paramagnetic spin state can be realized in the QSL state. 
This has been theoretically predicted as a paramagnetic state derived from the spinon Fermi surface \cite{Lee2005,Motrunich2005,Savary2017}, and such an example has actually been reported in EtMt$_3$Sb[Pd(dmit)$_2$]$_2$ \cite{Watanabe2012}. 
The second is that if the semiconducting behavior of \BEDSe\ is primarily due not to strong electronic correlations but to inhomogeneity arising from disorder in the cyano groups, the system can exhibit the frozen charge degree of freedom by disorder and the surviving spin degree of freedom. 
For reference, weak temperature dependence of $\chi_{\rm s}$ has been reported in systems where electrical conductivity is dominated by disorder \cite{Furukawa2015, Yasaka2024}. 

Here, we note that $\chi_{\rm s}$ of \BEDSe\ is larger than that of \ET, which is well described by a localized spin model. 
If \BEDSe\ lies in the insulating phase near the Mott boundary, this enhanced susceptibility may indicate an increase in the spinon Fermi surface contribution near the Mott transition \cite{Florens2002}. 
On the other hand, if \BEDSe\ is in the metallic phase close to the Mott transition, the enhancement could reflect critical growth of the spin susceptibility near the transition \cite{Dasari2017}. 
In either case, these results are expected to reveal the peculiar magnetic properties of a frustrated system near the Mott transition.

The inset in Fig.~\ref{sus} shows the magnetic field dependence of $\chi_{\rm s}$ at low temperatures.
At $0.05$~T, the magnetic susceptibility increases sharply below approximately $5$~K, and this increase is progressively suppressed as the magnetic field is increased.
Similar behavior has been discussed for \ET\ \cite{Isono2016, Riedl2019, Miksch2021}. 
Recent studies suggest that the understanding of magnetism in \ET\ is shifting from the previously proposed spin-liquid picture. 
The magnetism in this material is now thought to arise from the formation of spin singlets in a valence-bond-solid state, accompanied by the emergence of ``defect spins,'' which are unpaired localized spins believed to result from disorder effects \cite{Riedl2019, Miksch2021}. 
The increase in spin susceptibility at low temperatures, as well as its magnetic field dependence, is considered to be strongly influenced by such disorder.

\vspace{1\baselineskip}

Finally, we comment on the significance of \BEDSe\ salt for studies on QSL in organic systems.
Various physical properties have been measured on \ET, 
and the influence of the disordered  Cu$_2$(CN)$_3^-$ anion structure, as well as spin frustration, on the electronic properties has been pointed out \cite{Pinteric2014, Dressel2016}.
We believe that \BEDSe, in which only the effective pressure changes without altering the anion structure, will serve as a valuable reference material for clarifying the mysterious properties observed in \ET, such as the conflicting experimental results on low-energy spin excitations \cite{Yamashita2008,Yamashita2009}, the $6$~K anomaly \cite{Manna2010,Kobayashi2020,Matsuura2022}, valence bond solid formation \cite{Miksch2021}, and dielectric anomalies \cite{Abdel-Jawad2010}.
Furthermore, our measurements clarify how the spin susceptibility behaves on an almost ideal triangular lattice situated even closer to the Mott boundary. 
Because it is experimentally challenging to accurately measure static magnetic susceptibility under pressure in organic conductors, the present findings are particularly valuable. 
The obtained spin susceptibility of \BEDSe\ that cannot be reproduced by a simple localized triangular-lattice model serves as a valuable benchmark for theoretical studies on a frustrated Hubbard model that seek to capture correlation and frustration effects on the verge of a Mott transition.

\section{Summary}
We synthesized \BEDSe, which is revealed to be isostructural with a QSL candidate \ET, and investigated its electric and magnetic properties.
At ambient pressure, conductivities of \BEDSe\ and \ET\ can be explained by the same hopping mechanism. 
The estimated activation energies for \BEDSe\ at ambient pressure, along with the observed critical pressures for the insulator--metal and superconducting transitions, indicate that \BEDSe\ is located on the pressurized side, closer to the Mott boundary, compared to \ET. 
In contrast to $\chi_{\rm s}$ of \ET, which is explained by the localized spin model, 
\BEDSe\ exhibits a larger $\chi_{\rm s}$ with only weak temperature dependence.
The proximity to the Mott transition, resulting from a broader bandwidth of \BEDSe, can explain the enhanced Pauli paramagnetic-like susceptibility despite semiconducting transport.

\begin{acknowledgement}
The authors would like to thank A. Kawamoto and T. Yamamoto for their fruitful discussion. 
Single-crystal x-ray diffraction analyses were performed using a Bruker D8 QUEST ECO diffractometer installed at the Comprehensive Analysis Center for Science, Saitama University. 
This work was partially supported by the Japan Society for the Promotion of Science KAKENHI Grant Number 24K17002.
\end{acknowledgement}

\providecommand{\latin}[1]{#1}
\makeatletter
\providecommand{\doi}
  {\begingroup\let\do\@makeother\dospecials
  \catcode`\{=1 \catcode`\}=2 \doi@aux}
\providecommand{\doi@aux}[1]{\endgroup\texttt{#1}}
\makeatother
\providecommand*\mcitethebibliography{\thebibliography}
\csname @ifundefined\endcsname{endmcitethebibliography}
  {\let\endmcitethebibliography\endthebibliography}{}


\begin{mcitethebibliography}{77}
\providecommand*\natexlab[1]{#1}
\providecommand*\mciteSetBstSublistMode[1]{}
\providecommand*\mciteSetBstMaxWidthForm[2]{}
\providecommand*\mciteBstWouldAddEndPuncttrue
  {\def\EndOfBibitem{\unskip.}}
\providecommand*\mciteBstWouldAddEndPunctfalse
  {\let\EndOfBibitem\relax}
\providecommand*\mciteSetBstMidEndSepPunct[3]{}
\providecommand*\mciteSetBstSublistLabelBeginEnd[3]{}
\providecommand*\EndOfBibitem{}
\mciteSetBstSublistMode{f}
\mciteSetBstMaxWidthForm{subitem}{(\alph{mcitesubitemcount})}
\mciteSetBstSublistLabelBeginEnd
  {\mcitemaxwidthsubitemform\space}
  {\relax}
  {\relax}

\bibitem[Ramirez(1994)]{Ramirez1994}
Ramirez,~A.~P. {Strongly Geometrically Frustrated Magnets}. \emph{Annual Review
  of Materials Science} \textbf{1994}, \emph{24}, 453--480\relax
\mciteBstWouldAddEndPuncttrue
\mciteSetBstMidEndSepPunct{\mcitedefaultmidpunct}
{\mcitedefaultendpunct}{\mcitedefaultseppunct}\relax
\EndOfBibitem
\bibitem[Balents(2010)]{Balents2010}
Balents,~L. {Spin liquids in frustrated magnets}. \emph{Nature} \textbf{2010},
  \emph{464}, 199--208\relax
\mciteBstWouldAddEndPuncttrue
\mciteSetBstMidEndSepPunct{\mcitedefaultmidpunct}
{\mcitedefaultendpunct}{\mcitedefaultseppunct}\relax
\EndOfBibitem
\bibitem[Savary and Balents(2017)Savary, and Balents]{Savary2017}
Savary,~L.; Balents,~L. {Quantum spin liquids: a review}. \emph{Reports on
  Progress in Physics} \textbf{2017}, \emph{80}, 016502\relax
\mciteBstWouldAddEndPuncttrue
\mciteSetBstMidEndSepPunct{\mcitedefaultmidpunct}
{\mcitedefaultendpunct}{\mcitedefaultseppunct}\relax
\EndOfBibitem
\bibitem[Shimizu \latin{et~al.}(2003)Shimizu, Miyagawa, Kanoda, Maesato, and
  Saito]{Shimizu2003}
Shimizu,~Y.; Miyagawa,~K.; Kanoda,~K.; Maesato,~M.; Saito,~G. {Spin liquid
  state in an organic Mott insulator with a triangular lattice.} \emph{Physical
  Review Letters} \textbf{2003}, \emph{91}, 107001\relax
\mciteBstWouldAddEndPuncttrue
\mciteSetBstMidEndSepPunct{\mcitedefaultmidpunct}
{\mcitedefaultendpunct}{\mcitedefaultseppunct}\relax
\EndOfBibitem
\bibitem[Kagawa \latin{et~al.}(2013)Kagawa, Sato, Miyagawa, Kanoda, Tokura,
  Kobayashi, Kumai, and Murakami]{Kagawa2013}
Kagawa,~F.; Sato,~T.; Miyagawa,~K.; Kanoda,~K.; Tokura,~Y.; Kobayashi,~K.;
  Kumai,~R.; Murakami,~Y. {Charge-cluster glass in an organic conductor}.
  \emph{Nature Physics} \textbf{2013}, \emph{9}, 419--422\relax
\mciteBstWouldAddEndPuncttrue
\mciteSetBstMidEndSepPunct{\mcitedefaultmidpunct}
{\mcitedefaultendpunct}{\mcitedefaultseppunct}\relax
\EndOfBibitem
\bibitem[Sato \latin{et~al.}(2014)Sato, Kagawa, Kobayashi, Ueda, Mori,
  Miyagawa, Kanoda, Kumai, Murakami, and Tokura]{Sato2014}
Sato,~T.; Kagawa,~F.; Kobayashi,~K.; Ueda,~A.; Mori,~H.; Miyagawa,~K.;
  Kanoda,~K.; Kumai,~R.; Murakami,~Y.; Tokura,~Y. {Systematic variations in the
  charge-glass-forming ability of geometrically frustrated
  $\theta$-(BEDT-TTF)$_{2}X$ organic conductors}. \emph{Journal of the Physical
  Society of Japan} \textbf{2014}, \emph{83}, 083602\relax
\mciteBstWouldAddEndPuncttrue
\mciteSetBstMidEndSepPunct{\mcitedefaultmidpunct}
{\mcitedefaultendpunct}{\mcitedefaultseppunct}\relax
\EndOfBibitem
\bibitem[Hiramatsu \latin{et~al.}(2015)Hiramatsu, Yoshida, Saito, Otsuka,
  Yamochi, Maesato, Shimizu, Ito, and Kishida]{Hiramatsu2015}
Hiramatsu,~T.; Yoshida,~Y.; Saito,~G.; Otsuka,~A.; Yamochi,~H.; Maesato,~M.;
  Shimizu,~Y.; Ito,~H.; Kishida,~H. {Quantum spin liquid: design of a quantum
  spin liquid next to a superconducting state based on a dimer-type ET Mott
  insulator}. \emph{J. Mater. Chem. C} \textbf{2015}, \emph{3},
  1378--1388\relax
\mciteBstWouldAddEndPuncttrue
\mciteSetBstMidEndSepPunct{\mcitedefaultmidpunct}
{\mcitedefaultendpunct}{\mcitedefaultseppunct}\relax
\EndOfBibitem
\bibitem[Pustogow(2022)]{Pustogow2022}
Pustogow,~A. {Thirty-Year Anniversary of $\kappa$-(BEDT-TTF)$_2$Cu$_2$(CN)$_3$:
  Reconciling the Spin Gap in a Spin-Liquid Candidate}. \emph{Solids}
  \textbf{2022}, \emph{3}, 93--110\relax
\mciteBstWouldAddEndPuncttrue
\mciteSetBstMidEndSepPunct{\mcitedefaultmidpunct}
{\mcitedefaultendpunct}{\mcitedefaultseppunct}\relax
\EndOfBibitem
\bibitem[Yamashita \latin{et~al.}(2009)Yamashita, Nakata, Kasahara, Sasaki,
  Yoneyama, Kobayashi, Fujimoto, Shibauchi, and Matsuda]{Yamashita2009}
Yamashita,~M.; Nakata,~N.; Kasahara,~Y.; Sasaki,~T.; Yoneyama,~N.;
  Kobayashi,~N.; Fujimoto,~S.; Shibauchi,~T.; Matsuda,~Y. {Thermal-transport
  measurements in a quantum spin-liquid state of the frustrated triangular
  magnet $\kappa$-(BEDT-TTF)$_2$Cu$_2$(CN)$_3$}. \emph{Nature Physics}
  \textbf{2009}, \emph{5}, 44--47\relax
\mciteBstWouldAddEndPuncttrue
\mciteSetBstMidEndSepPunct{\mcitedefaultmidpunct}
{\mcitedefaultendpunct}{\mcitedefaultseppunct}\relax
\EndOfBibitem
\bibitem[Yamashita \latin{et~al.}(2008)Yamashita, Nakazawa, Oguni, Oshima,
  Nojiri, Shimizu, Miyagawa, and Kanoda]{Yamashita2008}
Yamashita,~S.; Nakazawa,~Y.; Oguni,~M.; Oshima,~Y.; Nojiri,~H.; Shimizu,~Y.;
  Miyagawa,~K.; Kanoda,~K. {Thermodynamic properties of a spin-1/2 spin-liquid
  state in a $\kappa$-type organic salt}. \emph{Nature Physics} \textbf{2008},
  \emph{4}, 459--462\relax
\mciteBstWouldAddEndPuncttrue
\mciteSetBstMidEndSepPunct{\mcitedefaultmidpunct}
{\mcitedefaultendpunct}{\mcitedefaultseppunct}\relax
\EndOfBibitem
\bibitem[Abdel-Jawad \latin{et~al.}(2010)Abdel-Jawad, Terasaki, Sasaki,
  Yoneyama, Kobayashi, Uesu, and Hotta]{Abdel-Jawad2010}
Abdel-Jawad,~M.; Terasaki,~I.; Sasaki,~T.; Yoneyama,~N.; Kobayashi,~N.;
  Uesu,~Y.; Hotta,~C. {Anomalous dielectric response in the dimer Mott
  insulator $\kappa$-(BEDT-TTF)$_2$Cu$_2$(CN)$_3$}. \emph{Physical Review B}
  \textbf{2010}, \emph{82}, 125119\relax
\mciteBstWouldAddEndPuncttrue
\mciteSetBstMidEndSepPunct{\mcitedefaultmidpunct}
{\mcitedefaultendpunct}{\mcitedefaultseppunct}\relax
\EndOfBibitem
\bibitem[Manna \latin{et~al.}(2010)Manna, de~Souza, Br{\"{u}}hl, Schlueter, and
  Lang]{Manna2010}
Manna,~R.~S.; de~Souza,~M.; Br{\"{u}}hl,~A.; Schlueter,~J.~A.; Lang,~M.
  {Lattice Effects and Entropy Release at the Low-Temperature Phase Transition
  in the Spin-Liquid Candidate $\kappa$-(BEDT-TTF)$_2$Cu$_2$(CN)$_3$}.
  \emph{Physical Review Letters} \textbf{2010}, \emph{104}, 016403\relax
\mciteBstWouldAddEndPuncttrue
\mciteSetBstMidEndSepPunct{\mcitedefaultmidpunct}
{\mcitedefaultendpunct}{\mcitedefaultseppunct}\relax
\EndOfBibitem
\bibitem[Kobayashi \latin{et~al.}(2020)Kobayashi, Ding, Taniguchi, Satoh,
  Kawamoto, and Furukawa]{Kobayashi2020}
Kobayashi,~T.; Ding,~Q.-P.; Taniguchi,~H.; Satoh,~K.; Kawamoto,~A.;
  Furukawa,~Y. {Charge disproportionation in the spin-liquid candidate
  $\kappa$-(ET)$_2$Cu$_2$(CN)$_3$ at $6$~K revealed by $^{63}$Cu NQR
  measurements}. \emph{Physical Review Research} \textbf{2020}, \emph{2},
  042023\relax
\mciteBstWouldAddEndPuncttrue
\mciteSetBstMidEndSepPunct{\mcitedefaultmidpunct}
{\mcitedefaultendpunct}{\mcitedefaultseppunct}\relax
\EndOfBibitem
\bibitem[Matsuura \latin{et~al.}(2022)Matsuura, Sasaki, Naka, M{\"{u}}ller,
  Stockert, Piovano, Yoneyama, and Lang]{Matsuura2022}
Matsuura,~M.; Sasaki,~T.; Naka,~M.; M{\"{u}}ller,~J.; Stockert,~O.;
  Piovano,~A.; Yoneyama,~N.; Lang,~M. {Phonon renormalization effects
  accompanying the $6$~K anomaly in the quantum spin liquid candidate
  $\kappa$-(BEDT-TTF)$_2$Cu$_2$(CN)$_3$}. \emph{Physical Review Research}
  \textbf{2022}, \emph{4}, L042047\relax
\mciteBstWouldAddEndPuncttrue
\mciteSetBstMidEndSepPunct{\mcitedefaultmidpunct}
{\mcitedefaultendpunct}{\mcitedefaultseppunct}\relax
\EndOfBibitem
\bibitem[Miksch \latin{et~al.}(2021)Miksch, Pustogow, Rahim, Bardin, Kanoda,
  Schlueter, H{\"{u}}bner, Scheffler, and Dressel]{Miksch2021}
Miksch,~B.; Pustogow,~A.; Rahim,~M.~J.; Bardin,~A.~A.; Kanoda,~K.;
  Schlueter,~J.~A.; H{\"{u}}bner,~R.; Scheffler,~M.; Dressel,~M. {Gapped
  magnetic ground state in quantum spin liquid candidate
  $\kappa$-(BEDT-TTF)$_2$Cu$_2$(CN)$_3$}. \emph{Science} \textbf{2021},
  \emph{372}, 276--279\relax
\mciteBstWouldAddEndPuncttrue
\mciteSetBstMidEndSepPunct{\mcitedefaultmidpunct}
{\mcitedefaultendpunct}{\mcitedefaultseppunct}\relax
\EndOfBibitem
\bibitem[Hiramatsu \latin{et~al.}(2017)Hiramatsu, Yoshida, Saito, Otsuka,
  Yamochi, Maesato, Shimizu, Ito, Nakamura, Kishida, Watanabe, and
  Kumai]{Hiramatsu2017}
Hiramatsu,~T.; Yoshida,~Y.; Saito,~G.; Otsuka,~A.; Yamochi,~H.; Maesato,~M.;
  Shimizu,~Y.; Ito,~H.; Nakamura,~Y.; Kishida,~H.; Watanabe,~M.; Kumai,~R.
  \emph{Bulletin of the Chemical Society of Japan} \textbf{2017}, \emph{90},
  1073--1082\relax
\mciteBstWouldAddEndPuncttrue
\mciteSetBstMidEndSepPunct{\mcitedefaultmidpunct}
{\mcitedefaultendpunct}{\mcitedefaultseppunct}\relax
\EndOfBibitem
\bibitem[Yoshida \latin{et~al.}(2019)Yoshida, Maesato, Tomeno, Kimura, Saito,
  Nakamura, Kishida, and Kitagawa]{Yoshida2019a}
Yoshida,~Y.; Maesato,~M.; Tomeno,~S.; Kimura,~Y.; Saito,~G.; Nakamura,~Y.;
  Kishida,~H.; Kitagawa,~H. {Partial Substitution of Ag(I) for Cu(I) in Quantum
  Spin Liquid $\kappa$-(ET)$_2$Cu$_2$(CN)$_3$, Where ET Is
  Bis(ethylenedithio)tetrathiafulvalene}. \emph{Inorganic Chemistry}
  \textbf{2019}, \emph{58}, 4820--4827\relax
\mciteBstWouldAddEndPuncttrue
\mciteSetBstMidEndSepPunct{\mcitedefaultmidpunct}
{\mcitedefaultendpunct}{\mcitedefaultseppunct}\relax
\EndOfBibitem
\bibitem[Tomeno \latin{et~al.}(2020)Tomeno, Maesato, Yoshida, Kiswandhi, and
  Kitagawa]{Tomeno2020}
Tomeno,~S.; Maesato,~M.; Yoshida,~Y.; Kiswandhi,~A.; Kitagawa,~H.
  {Triangular-Lattice Organic Mott Insulator with a Disorder-Free Polyanion}.
  \emph{Inorganic Chemistry} \textbf{2020}, \emph{59}, 8647--8651\relax
\mciteBstWouldAddEndPuncttrue
\mciteSetBstMidEndSepPunct{\mcitedefaultmidpunct}
{\mcitedefaultendpunct}{\mcitedefaultseppunct}\relax
\EndOfBibitem
\bibitem[Pinteri{\'{c}} \latin{et~al.}(2014)Pinteri{\'{c}}, {\v{C}}ulo, Milat,
  Basleti{\'{c}}, Korin-Hamzi{\'{c}}, Tafra, Hamzi{\'{c}}, Ivek, Peterseim,
  Miyagawa, Kanoda, Schlueter, Dressel, and Tomi{\'{c}}]{Pinteric2014}
Pinteri{\'{c}},~M.; {\v{C}}ulo,~M.; Milat,~O.; Basleti{\'{c}},~M.;
  Korin-Hamzi{\'{c}},~B.; Tafra,~E.; Hamzi{\'{c}},~A.; Ivek,~T.; Peterseim,~T.;
  Miyagawa,~K.; Kanoda,~K.; Schlueter,~J.~A.; Dressel,~M.; Tomi{\'{c}},~S.
  {Anisotropic charge dynamics in the quantum spin-liquid candidate
  $\kappa$-(BEDT-TTF)$_2$Cu$_2$(CN)$_3$}. \emph{Physical Review B}
  \textbf{2014}, \emph{90}, 195139\relax
\mciteBstWouldAddEndPuncttrue
\mciteSetBstMidEndSepPunct{\mcitedefaultmidpunct}
{\mcitedefaultendpunct}{\mcitedefaultseppunct}\relax
\EndOfBibitem
\bibitem[Dressel \latin{et~al.}(2016)Dressel, Lazi^^c4^^87, Pustogow, Zhukova,
  Gorshunov, Schlueter, Milat, Gumhalter, and Tomi^^c4^^87]{Dressel2016}
Dressel,~M.; Lazi^^c4^^87,~P.; Pustogow,~A.; Zhukova,~E.; Gorshunov,~B.;
  Schlueter,~J.~A.; Milat,~O.; Gumhalter,~B.; Tomi^^c4^^87,~S. Lattice
  vibrations of the charge-transfer salt $\kappa$-(BEDT-TTF)$_2$Cu$_2$(CN)$_3$
  : Comprehensive explanation of the electrodynamic response in a spin-liquid
  compound. \emph{Physical Review B} \textbf{2016}, \emph{93}, 081201\relax
\mciteBstWouldAddEndPuncttrue
\mciteSetBstMidEndSepPunct{\mcitedefaultmidpunct}
{\mcitedefaultendpunct}{\mcitedefaultseppunct}\relax
\EndOfBibitem
\bibitem[Shimizu \latin{et~al.}(2003)Shimizu, Maesato, Saito, Drozdova, and
  Ouahab]{Shimizu2003a}
Shimizu,~Y.; Maesato,~M.; Saito,~G.; Drozdova,~O.; Ouahab,~L. {Transport
  properties of a Mott insulator $\kappa$-(ET)$_2$Cu$_2$(CN)$_3$ under the
  uniaxial strain}. \emph{Synthetic Metals} \textbf{2003}, \emph{133-134},
  225--226\relax
\mciteBstWouldAddEndPuncttrue
\mciteSetBstMidEndSepPunct{\mcitedefaultmidpunct}
{\mcitedefaultendpunct}{\mcitedefaultseppunct}\relax
\EndOfBibitem
\bibitem[Pustogow \latin{et~al.}(2021)Pustogow, Saito, L{\"{o}}hle, {Sanz
  Alonso}, Kawamoto, Dobrosavljevi{\'{c}}, Dressel, and Fratini]{Pustogow2021}
Pustogow,~A.; Saito,~Y.; L{\"{o}}hle,~A.; {Sanz Alonso},~M.; Kawamoto,~A.;
  Dobrosavljevi{\'{c}},~V.; Dressel,~M.; Fratini,~S. {Rise and fall of Landau's
  quasiparticles while approaching the Mott transition}. \emph{Nature
  Communications} \textbf{2021}, \emph{12}, 1--8\relax
\mciteBstWouldAddEndPuncttrue
\mciteSetBstMidEndSepPunct{\mcitedefaultmidpunct}
{\mcitedefaultendpunct}{\mcitedefaultseppunct}\relax
\EndOfBibitem
\bibitem[Saito \latin{et~al.}(2021)Saito, L{\"{o}}hle, Kawamoto, Pustogow, and
  Dressel]{Saito2021}
Saito,~Y.; L{\"{o}}hle,~A.; Kawamoto,~A.; Pustogow,~A.; Dressel,~M.
  {Pressure-Tuned Superconducting Dome in Chemically-Substituted
  $\kappa$-(BEDT-TTF)$_2$Cu$_2$(CN)$_3$}. \emph{Crystals} \textbf{2021},
  \emph{11}, 817\relax
\mciteBstWouldAddEndPuncttrue
\mciteSetBstMidEndSepPunct{\mcitedefaultmidpunct}
{\mcitedefaultendpunct}{\mcitedefaultseppunct}\relax
\EndOfBibitem
\bibitem[Mori \latin{et~al.}(1984)Mori, Kobayashi, Sasaki, Kobayashi, Saito,
  and Inokuchi]{Mori1984}
Mori,~T.; Kobayashi,~A.; Sasaki,~Y.; Kobayashi,~H.; Saito,~G.; Inokuchi,~H.
  {The Intermolecular Interaction of Tetrathiafulvalene and
  Bis(ethylenedithio)tetrathiafulvalene in Organic Metals. Calculation of
  Orbital Overlaps and Models of Energy-band Structures}. \emph{Bulletin of the
  Chemical Society of Japan} \textbf{1984}, \emph{57}, 627--633\relax
\mciteBstWouldAddEndPuncttrue
\mciteSetBstMidEndSepPunct{\mcitedefaultmidpunct}
{\mcitedefaultendpunct}{\mcitedefaultseppunct}\relax
\EndOfBibitem
\bibitem[Sakata \latin{et~al.}(1998)Sakata, Sato, Miyazaki, Enoki, Okano, and
  Kato]{Sakata1998}
Sakata,~J.-i.; Sato,~H.; Miyazaki,~A.; Enoki,~T.; Okano,~Y.; Kato,~R.
  {Superconductivity in new organic conductor
  $\kappa$-(BEDSe-TTF)$_2$CuN(CN)$_2$Br}. \emph{Solid State Communications}
  \textbf{1998}, \emph{108}, 377--381\relax
\mciteBstWouldAddEndPuncttrue
\mciteSetBstMidEndSepPunct{\mcitedefaultmidpunct}
{\mcitedefaultendpunct}{\mcitedefaultseppunct}\relax
\EndOfBibitem
\bibitem[Imajo \latin{et~al.}(2022)Imajo, Kato, Marckwardt, Yesil, Akutsu, and
  Nakazawa]{Imajo2022}
Imajo,~S.; Kato,~N.; Marckwardt,~R.~J.; Yesil,~E.; Akutsu,~H.; Nakazawa,~Y.
  {Persistence of fermionic spin excitations through a genuine Mott transition
  in $\kappa$-type organics}. \emph{Physical Review B} \textbf{2022},
  \emph{105}, 125130\relax
\mciteBstWouldAddEndPuncttrue
\mciteSetBstMidEndSepPunct{\mcitedefaultmidpunct}
{\mcitedefaultendpunct}{\mcitedefaultseppunct}\relax
\EndOfBibitem
\bibitem[{Bruker}(2022)]{bruker2022apex5}
{Bruker}, APEX5. Bruker AXS Inc.: Madison, Wisconsin, USA, 2022\relax
\mciteBstWouldAddEndPuncttrue
\mciteSetBstMidEndSepPunct{\mcitedefaultmidpunct}
{\mcitedefaultendpunct}{\mcitedefaultseppunct}\relax
\EndOfBibitem
\bibitem[Sheldrick(2015)]{Sheldrick2015a}
Sheldrick,~G.~M. {SHELXT ^^e2^^80^^93 Integrated space-group and
  crystal-structure determination}. \emph{Acta Crystallographica Section A
  Foundations and Advances} \textbf{2015}, \emph{71}, 3--8\relax
\mciteBstWouldAddEndPuncttrue
\mciteSetBstMidEndSepPunct{\mcitedefaultmidpunct}
{\mcitedefaultendpunct}{\mcitedefaultseppunct}\relax
\EndOfBibitem
\bibitem[Sheldrick(2015)]{Sheldrick2015}
Sheldrick,~G.~M. {Crystal structure refinement with SHELXL}. \emph{Acta
  Crystallographica Section C Structural Chemistry} \textbf{2015}, \emph{71},
  3--8\relax
\mciteBstWouldAddEndPuncttrue
\mciteSetBstMidEndSepPunct{\mcitedefaultmidpunct}
{\mcitedefaultendpunct}{\mcitedefaultseppunct}\relax
\EndOfBibitem
\bibitem[Dolomanov \latin{et~al.}(2009)Dolomanov, Bourhis, Gildea, Howard, and
  Puschmann]{Dolomanov2009}
Dolomanov,~O.~V.; Bourhis,~L.~J.; Gildea,~R.~J.; Howard,~J. A.~K.;
  Puschmann,~H. {OLEX2 : a complete structure solution, refinement and analysis
  program}. \emph{Journal of Applied Crystallography} \textbf{2009}, \emph{42},
  339--341\relax
\mciteBstWouldAddEndPuncttrue
\mciteSetBstMidEndSepPunct{\mcitedefaultmidpunct}
{\mcitedefaultendpunct}{\mcitedefaultseppunct}\relax
\EndOfBibitem
\bibitem[Cry()]{Crystallographic}
Crystal data for $\kappa$-(BEST)$_2$Cu$_2$(CN)$_3$
  (C$_{23}$H$_{16}$N$_3$S$_{8}$Cu$_{2}$Se$_{8}$, $M = 1349.63$): $T = 300$~K;
  monoclinic space group $P2_1/c$, $a = 16.4504(6)$~\AA, $b = 8.6983(3)$~\AA,
  $c = 13.5492(5)$~\AA, $\beta = 113.9090(10)^\circ$, $V = 1772.4(1)$~\AA$^3$,
  $Z = 2$, $d_{\mathrm{calc}} = 2.529$~g~cm$^{-3}$, $\mu$(Mo K$\alpha$) $=
  9.911$~mm$^{-1}$, $F(000) = 1266$, 6141 reflections used, 219 refined
  parameters, $R_1 = 0.0371$ [for $I > 2\sigma(I)$], $wR_2 = 0.0891$ (for all
  data), GOF = 1.017, BASF = 0.0175(3); CCDC 2442311. $T = 150$~K; monoclinic
  space group $P2_1/c$, $a = 16.4171(11)$~\AA, $b = 8.6876(6)$~\AA, $c =
  13.4152(9)$~\AA, $\beta = 114.769(2)^\circ$, $V = 1737.3(2)$~\AA$^3$, $Z =
  2$, $d_{\mathrm{calc}} = 2.580$~g~cm$^{-3}$, $\mu$(Mo K$\alpha$) $=
  10.111$~mm$^{-1}$, $F(000) = 1266$, 9233 reflections used, 195 refined
  parameters, $R_1 = 0.0350$ [for $I > 2\sigma(I)$], $wR_2 = 0.0831$ (for all
  data), GOF = 1.029, BASF = 0.00716(18); CCDC 2475419.\relax
\mciteBstWouldAddEndPunctfalse
\mciteSetBstMidEndSepPunct{\mcitedefaultmidpunct}
{}{\mcitedefaultseppunct}\relax
\EndOfBibitem
\bibitem[Komatsu \latin{et~al.}(1991)Komatsu, Nakamura, Matsukawa, Yamochi,
  Saito, Ito, Ishiguro, Kusunoki, and Sakaguchi]{Komatsu1991}
Komatsu,~T.; Nakamura,~T.; Matsukawa,~N.; Yamochi,~H.; Saito,~G.; Ito,~H.;
  Ishiguro,~T.; Kusunoki,~M.; Sakaguchi,~K.-i. {New ambient-pressure organic
  superconductors based on BEDT-TTF, Cu, N(CN)$_2$ and CN with $T_c$ = $10.7$~K
  and $3.8$~K}. \emph{Solid State Communications} \textbf{1991}, \emph{80},
  843--847\relax
\mciteBstWouldAddEndPuncttrue
\mciteSetBstMidEndSepPunct{\mcitedefaultmidpunct}
{\mcitedefaultendpunct}{\mcitedefaultseppunct}\relax
\EndOfBibitem
\bibitem[Geiser \latin{et~al.}(1991)Geiser, Wang, Carlson, Williams, Charlier,
  Heindl, Yaconi, Love, Lathrop, Schirber, Overmyer, Ren, and
  Whangbo]{Geiser1991a}
Geiser,~U.; Wang,~H.~H.; Carlson,~K.~D.; Williams,~J.~M.; Charlier,~H.~A.;
  Heindl,~J.~E.; Yaconi,~G.~A.; Love,~B.~J.; Lathrop,~M.~W.; Schirber,~J.~E.;
  Overmyer,~D.~L.; Ren,~J.; Whangbo,~M.-H. {Superconductivity at $2.8$~K and
  $1.5$~kbar in $\kappa$-(BEDT-TTF)$_2$Cu$_2$(CN)$_3$: the first organic
  superconductor containing a polymeric copper cyanide anion}. \emph{Inorganic
  Chemistry} \textbf{1991}, \emph{30}, 2586--2588\relax
\mciteBstWouldAddEndPuncttrue
\mciteSetBstMidEndSepPunct{\mcitedefaultmidpunct}
{\mcitedefaultendpunct}{\mcitedefaultseppunct}\relax
\EndOfBibitem
\bibitem[Huc()]{Huckelparameter}
The $ \zeta$ exponent (ionization potential (eV)) for atomic orbitals are as
  follows. Se : 4d 2.44 (-20.0), 4p 2.07 (-10.8), 4s 1.5 (-5.44). S : 3d 2.122
  (-20.0), 3p 1.827 (-11.0), 3s 1.5 (-5.44). C : 2p 1.625 (-21.4), 2s 1.625
  (-11.4). H : 1s 1.3 (-13.6). These values were taken from those used in the
  literatures \cite{Mori2002,Ito2022}.\relax
\mciteBstWouldAddEndPunctfalse
\mciteSetBstMidEndSepPunct{\mcitedefaultmidpunct}
{}{\mcitedefaultseppunct}\relax
\EndOfBibitem
\bibitem[Uwatoko \latin{et~al.}(2002)Uwatoko, Todo, Ueda, Uchida, Kosaka, Mori,
  and Matsumoto]{Uwatoko2002}
Uwatoko,~Y.; Todo,~S.; Ueda,~K.; Uchida,~A.; Kosaka,~M.; Mori,~N.;
  Matsumoto,~T. {Material properties of Ni--Cr--Al alloy and design of a 4 GPa
  class non-magnetic high-pressure cell}. \emph{Journal of Physics: Condensed
  Matter} \textbf{2002}, \emph{14}, 11291--11296\relax
\mciteBstWouldAddEndPuncttrue
\mciteSetBstMidEndSepPunct{\mcitedefaultmidpunct}
{\mcitedefaultendpunct}{\mcitedefaultseppunct}\relax
\EndOfBibitem
\bibitem[Taniguchi \latin{et~al.}(2010)Taniguchi, Takeda, Satoh, Taniguchi,
  Komatsu, and Satoh]{Taniguchi2010}
Taniguchi,~H.; Takeda,~S.; Satoh,~R.; Taniguchi,~A.; Komatsu,~H.; Satoh,~K.
  {Short piston-cylinder pressure cells based on Ni--Cr--Al cylinders and their
  application to fragile materials}. \emph{Review of Scientific Instruments}
  \textbf{2010}, \emph{81}\relax
\mciteBstWouldAddEndPuncttrue
\mciteSetBstMidEndSepPunct{\mcitedefaultmidpunct}
{\mcitedefaultendpunct}{\mcitedefaultseppunct}\relax
\EndOfBibitem
\bibitem[Murata \latin{et~al.}(1997)Murata, Yoshino, Yadav, Honda, and
  Shirakawa]{Murata1997}
Murata,~K.; Yoshino,~H.; Yadav,~H.~O.; Honda,~Y.; Shirakawa,~N. {Pt resistor
  thermometry and pressure calibration in a clamped pressure cell with the
  medium, Daphne 7373}. \emph{Review of Scientific Instruments} \textbf{1997},
  \emph{68}, 2490--2493\relax
\mciteBstWouldAddEndPuncttrue
\mciteSetBstMidEndSepPunct{\mcitedefaultmidpunct}
{\mcitedefaultendpunct}{\mcitedefaultseppunct}\relax
\EndOfBibitem
\bibitem[Eiling and Schilling(1981)Eiling, and Schilling]{Eiling1981}
Eiling,~A.; Schilling,~J.~S. {Pressure and temperature dependence of electrical
  resistivity of Pb and Sn from 1-300K and 0-10 GPa-use as continuous resistive
  pressure monitor accurate over wide temperature range; superconductivity
  under pressure in Pb, Sn and In}. \emph{Journal of Physics F: Metal Physics}
  \textbf{1981}, \emph{11}, 623--639\relax
\mciteBstWouldAddEndPuncttrue
\mciteSetBstMidEndSepPunct{\mcitedefaultmidpunct}
{\mcitedefaultendpunct}{\mcitedefaultseppunct}\relax
\EndOfBibitem
\bibitem[Spa()]{Spacegroup}
Detailed structural analysis for $\kappa$-ET-CN has reported that the space
  group is $P\overline{1}$ even at room temperature \cite{Foury-Leylekian2018},
  but its structure determination requires the capture of weak intensity
  reflections due to weak symmetry breaking, which is difficult to achieve with
  standard collection and refinement procedures. Therefore, we applied the
  previously reported space group $P2_{1}/c$ for both $\kappa$-ET-CN and
  $\kappa$-BEST-CN.\relax
\mciteBstWouldAddEndPunctfalse
\mciteSetBstMidEndSepPunct{\mcitedefaultmidpunct}
{}{\mcitedefaultseppunct}\relax
\EndOfBibitem
\bibitem[Jeschke \latin{et~al.}(2012)Jeschke, de~Souza, Valent{\'{i}}, Manna,
  Lang, and Schlueter]{Jeschke2012}
Jeschke,~H.~O.; de~Souza,~M.; Valent{\'{i}},~R.; Manna,~R.~S.; Lang,~M.;
  Schlueter,~J.~A. {Temperature dependence of structural and electronic
  properties of the spin-liquid candidate
  $\kappa$-(BEDT-TTF)$_2$Cu$_2$(CN)$_3$}. \emph{Physical Review B}
  \textbf{2012}, \emph{85}, 035125\relax
\mciteBstWouldAddEndPuncttrue
\mciteSetBstMidEndSepPunct{\mcitedefaultmidpunct}
{\mcitedefaultendpunct}{\mcitedefaultseppunct}\relax
\EndOfBibitem
\bibitem[Komatsu \latin{et~al.}(1996)Komatsu, Matsukawa, Inoue, and
  Saito]{Komatsu1996}
Komatsu,~T.; Matsukawa,~N.; Inoue,~T.; Saito,~G. {Realization of
  Superconductivity at Ambient Pressure by Band-Filling Control in
  $\kappa$-(BEDT-TTF)$_2$Cu$_2$(CN)$_3$}. \emph{Journal of the Physical Society
  of Japan} \textbf{1996}, \emph{65}, 1340--1354\relax
\mciteBstWouldAddEndPuncttrue
\mciteSetBstMidEndSepPunct{\mcitedefaultmidpunct}
{\mcitedefaultendpunct}{\mcitedefaultseppunct}\relax
\EndOfBibitem
\bibitem[Kanoda(1997)]{Kanoda1997c}
Kanoda,~K. {Recent progress in NMR studies on organic conductors}.
  \emph{Hyperfine Interactions} \textbf{1997}, \emph{104}, 235--249\relax
\mciteBstWouldAddEndPuncttrue
\mciteSetBstMidEndSepPunct{\mcitedefaultmidpunct}
{\mcitedefaultendpunct}{\mcitedefaultseppunct}\relax
\EndOfBibitem
\bibitem[Kawamoto \latin{et~al.}(2004)Kawamoto, Honma, and
  Kumagai]{Kawamoto2004a}
Kawamoto,~A.; Honma,~Y.; Kumagai,~K.-i. {Electron localization in the strongly
  correlated organic system $\kappa$-(BEDT-TTF)$_2X$ probed with nuclear
  magnetic resonance $^{13}$C-NMR}. \emph{Physical Review B} \textbf{2004},
  \emph{70}, 060510\relax
\mciteBstWouldAddEndPuncttrue
\mciteSetBstMidEndSepPunct{\mcitedefaultmidpunct}
{\mcitedefaultendpunct}{\mcitedefaultseppunct}\relax
\EndOfBibitem
\bibitem[Saito \latin{et~al.}(2018)Saito, Minamidate, Kawamoto, Matsunaga, and
  Nomura]{Saito2018}
Saito,~Y.; Minamidate,~T.; Kawamoto,~A.; Matsunaga,~N.; Nomura,~K.
  {Site-specific $^{13}$C NMR study on the locally distorted triangular lattice
  of the organic conductor $\kappa$-(BEDT-TTF)$_2$Cu$_2$(CN)$_3$}.
  \emph{Physical Review B} \textbf{2018}, \emph{98}, 205141\relax
\mciteBstWouldAddEndPuncttrue
\mciteSetBstMidEndSepPunct{\mcitedefaultmidpunct}
{\mcitedefaultendpunct}{\mcitedefaultseppunct}\relax
\EndOfBibitem
\bibitem[{\v{C}}ulo \latin{et~al.}(2019){\v{C}}ulo, Tafra, Mihaljevi{\'{c}},
  Basleti{\'{c}}, Kuve{\v{z}}di{\'{c}}, Ivek, Hamzi{\'{c}}, Tomi{\'{c}},
  Hiramatsu, Yoshida, Saito, Schlueter, Dressel, and
  Korin-Hamzi{\'{c}}]{Culo2019}
{\v{C}}ulo,~M.; Tafra,~E.; Mihaljevi{\'{c}},~B.; Basleti{\'{c}},~M.;
  Kuve{\v{z}}di{\'{c}},~M.; Ivek,~T.; Hamzi{\'{c}},~A.; Tomi{\'{c}},~S.;
  Hiramatsu,~T.; Yoshida,~Y.; Saito,~G.; Schlueter,~J.~A.; Dressel,~M.;
  Korin-Hamzi{\'{c}},~B. {Hall effect study of the $\kappa$-(ET)$_{2}X$ family:
  Evidence for Mott-Anderson localization}. \emph{Physical Review B}
  \textbf{2019}, \emph{99}, 045114\relax
\mciteBstWouldAddEndPuncttrue
\mciteSetBstMidEndSepPunct{\mcitedefaultmidpunct}
{\mcitedefaultendpunct}{\mcitedefaultseppunct}\relax
\EndOfBibitem
\bibitem[Byczuk \latin{et~al.}(2005)Byczuk, Hofstetter, and
  Vollhardt]{Byczuk2005}
Byczuk,~K.; Hofstetter,~W.; Vollhardt,~D. Mott-Hubbard Transition versus
  Anderson Localization in Correlated Electron Systems with Disorder.
  \emph{Physical Review Letters} \textbf{2005}, \emph{94}, 056404\relax
\mciteBstWouldAddEndPuncttrue
\mciteSetBstMidEndSepPunct{\mcitedefaultmidpunct}
{\mcitedefaultendpunct}{\mcitedefaultseppunct}\relax
\EndOfBibitem
\bibitem[Furukawa \latin{et~al.}(2018)Furukawa, Kobashi, Kurosaki, Miyagawa,
  and Kanoda]{Furukawa2018}
Furukawa,~T.; Kobashi,~K.; Kurosaki,~Y.; Miyagawa,~K.; Kanoda,~K.
  {Quasi-continuous transition from a Fermi liquid to a spin liquid in
  $\kappa$-(ET)$_2$Cu$_2$(CN)$_3$}. \emph{Nature Communications} \textbf{2018},
  \emph{9}, 307\relax
\mciteBstWouldAddEndPuncttrue
\mciteSetBstMidEndSepPunct{\mcitedefaultmidpunct}
{\mcitedefaultendpunct}{\mcitedefaultseppunct}\relax
\EndOfBibitem
\bibitem[Kurosaki \latin{et~al.}(2005)Kurosaki, Shimizu, Miyagawa, Kanoda, and
  Saito]{Kurosaki2005}
Kurosaki,~Y.; Shimizu,~Y.; Miyagawa,~K.; Kanoda,~K.; Saito,~G. {Mott transition
  from a spin liquid to a Fermi liquid in the spin-frustrated organic conductor
  $\kappa$-(ET)$_2$Cu$_2$(CN)$_3$}. \emph{Physical Review Letters}
  \textbf{2005}, \emph{95}, 177001\relax
\mciteBstWouldAddEndPuncttrue
\mciteSetBstMidEndSepPunct{\mcitedefaultmidpunct}
{\mcitedefaultendpunct}{\mcitedefaultseppunct}\relax
\EndOfBibitem
\bibitem[Murata \latin{et~al.}(1990)Murata, Ishibashi, Honda, Fortune,
  Tokumoto, Kinoshita, and Anzai]{Murata1990}
Murata,~K.; Ishibashi,~M.; Honda,~Y.; Fortune,~N.~A.; Tokumoto,~M.;
  Kinoshita,~N.; Anzai,~H. {Temperature dependence of hall effect in
  $\kappa$-(BEDT-TTF)$_2$Cu(NCS)$_2$}. \emph{Solid State Communications}
  \textbf{1990}, \emph{76}, 377--381\relax
\mciteBstWouldAddEndPuncttrue
\mciteSetBstMidEndSepPunct{\mcitedefaultmidpunct}
{\mcitedefaultendpunct}{\mcitedefaultseppunct}\relax
\EndOfBibitem
\bibitem[Taniguchi \latin{et~al.}(1999)Taniguchi, Kawamoto, and
  Kanoda]{Taniguchi1999}
Taniguchi,~H.; Kawamoto,~A.; Kanoda,~K. {Superconductor-insulator phase
  transformation of partially deuterated $\kappa$-(BEDT-TTF)$_2$Cu[N(CN)$_2$]Br
  by control of the cooling rate}. \emph{Physical Review B} \textbf{1999},
  \emph{59}, 8424--8427\relax
\mciteBstWouldAddEndPuncttrue
\mciteSetBstMidEndSepPunct{\mcitedefaultmidpunct}
{\mcitedefaultendpunct}{\mcitedefaultseppunct}\relax
\EndOfBibitem
\bibitem[Strack \latin{et~al.}(2005)Strack, Akinci, Paschenko, Wolf, Uhrig,
  Assmus, Lang, Schreuer, Wiehl, Schlueter, Wosnitza, Schweitzer, M{\"{u}}ller,
  and Wykhoff]{Strack2005}
Strack,~C.; Akinci,~C.; Paschenko,~V.; Wolf,~B.; Uhrig,~E.; Assmus,~W.;
  Lang,~M.; Schreuer,~J.; Wiehl,~L.; Schlueter,~J.~A.; Wosnitza,~J.;
  Schweitzer,~D.; M{\"{u}}ller,~J.; Wykhoff,~J. {Resistivity studies under
  hydrostatic pressure on a low-resistance variant of the quasi-two-dimensional
  organic superconductor $\kappa$-(BEDT-TTF)$_2$Cu[N(CN)$_2$]Br: Search for
  intrinsic scattering contributions}. \emph{Physical Review B} \textbf{2005},
  \emph{72}, 054511\relax
\mciteBstWouldAddEndPuncttrue
\mciteSetBstMidEndSepPunct{\mcitedefaultmidpunct}
{\mcitedefaultendpunct}{\mcitedefaultseppunct}\relax
\EndOfBibitem
\bibitem[Kobayashi and Kawamoto(2017)Kobayashi, and Kawamoto]{Kobayashi2017}
Kobayashi,~T.; Kawamoto,~A. {Evidence of antiferromagnetic fluctuation in the
  unconventional superconductor $\lambda$-(BETS)$_2$GaCl$_4$ by $^{13}$C NMR}.
  \emph{Physical Review B} \textbf{2017}, \emph{96}, 125115\relax
\mciteBstWouldAddEndPuncttrue
\mciteSetBstMidEndSepPunct{\mcitedefaultmidpunct}
{\mcitedefaultendpunct}{\mcitedefaultseppunct}\relax
\EndOfBibitem
\bibitem[Batsanov(2001)]{Batsanov2001}
Batsanov,~S.~S. {Van der Waals Radii of Elements}. \emph{Inorganic Materials}
  \textbf{2001}, \emph{37}, 871--885\relax
\mciteBstWouldAddEndPuncttrue
\mciteSetBstMidEndSepPunct{\mcitedefaultmidpunct}
{\mcitedefaultendpunct}{\mcitedefaultseppunct}\relax
\EndOfBibitem
\bibitem[Kobayashi \latin{et~al.}(2004)Kobayashi, Cui, and
  Kobayashi]{Kobayashi2004}
Kobayashi,~H.; Cui,~H.; Kobayashi,~A. Organic metals and superconductors based
  on BETS (BETS = Bis(ethylenedithio)tetraselenafulvalene). \emph{Chemical
  Reviews} \textbf{2004}, \emph{104}, 5265--5288\relax
\mciteBstWouldAddEndPuncttrue
\mciteSetBstMidEndSepPunct{\mcitedefaultmidpunct}
{\mcitedefaultendpunct}{\mcitedefaultseppunct}\relax
\EndOfBibitem
\bibitem[Courcet \latin{et~al.}(1998)Courcet, Malfant, Pokhodnia, and
  Cassoux]{Courcet1998}
Courcet,~T.; Malfant,~I.; Pokhodnia,~K.; Cassoux,~P.
  Bis(ethylenedithio)tetraselenafulvalene: short-cut synthesis, X-ray crystal
  structure and $\pi$-electron density distribution. \emph{New Journal of
  Chemistry} \textbf{1998}, \emph{22}, 585--589\relax
\mciteBstWouldAddEndPuncttrue
\mciteSetBstMidEndSepPunct{\mcitedefaultmidpunct}
{\mcitedefaultendpunct}{\mcitedefaultseppunct}\relax
\EndOfBibitem
\bibitem[Wang \latin{et~al.}(1989)Wang, Montgomery, Geiser, Porter, Carlson,
  Ferraro, Williams, Cariss, and Rubinstein]{Wang1989}
Wang,~H.~H.; Montgomery,~L.~K.; Geiser,~U.; Porter,~L.~C.; Carlson,~K.~D.;
  Ferraro,~J.~R.; Williams,~J.~M.; Cariss,~C.~S.; Rubinstein,~R.~L. Syntheses,
  structures, selected physical properties and band electronic structures of
  the bis(ethylenediseleno)tetrathiafulvalene salts, (BEDSe-TTF)$_2X$, X- =
  I$_3^-$, AuI$_2^-$, and IBr$_2^-$. \emph{Chemistry of Materials}
  \textbf{1989}, \emph{1}, 140--148\relax
\mciteBstWouldAddEndPuncttrue
\mciteSetBstMidEndSepPunct{\mcitedefaultmidpunct}
{\mcitedefaultendpunct}{\mcitedefaultseppunct}\relax
\EndOfBibitem
\bibitem[Lyubovskaya \latin{et~al.}(1997)Lyubovskaya, Zhilyaeva, Torunova,
  Bogdanova, Konovalikhin, Dyachenko, and Lyubovskii]{Lyubovskaya1997}
Lyubovskaya,~R.~N.; Zhilyaeva,~E.~I.; Torunova,~S.~A.; Bogdanova,~O.~A.;
  Konovalikhin,~S.~V.; Dyachenko,~O.~A.; Lyubovskii,~R.~B. BEDT-TTF, BEDO-TTF,
  and BEDSe-TTF salts with metal containing anions. \emph{Synthetic Metals}
  \textbf{1997}, \emph{85}, 1581--1582\relax
\mciteBstWouldAddEndPuncttrue
\mciteSetBstMidEndSepPunct{\mcitedefaultmidpunct}
{\mcitedefaultendpunct}{\mcitedefaultseppunct}\relax
\EndOfBibitem
\bibitem[Clemente-Le^^c3^^b3n \latin{et~al.}(2002)Clemente-Le^^c3^^b3n,
  Coronado, Gal^^c3^^a1n-Mascar^^c3^^b3s, Gim^^c3^^a9nez-Saiz,
  G^^c3^^b3mez-Garc^^c3^^ada, Fabre, Mousdis, and
  Papavassiliou]{Clemente-Leon2002}
Clemente-Le^^c3^^b3n,~M.; Coronado,~E.; Gal^^c3^^a1n-Mascar^^c3^^b3s,~J.~R.;
  Gim^^c3^^a9nez-Saiz,~C.; G^^c3^^b3mez-Garc^^c3^^ada,~C.~J.; Fabre,~J.~M.;
  Mousdis,~G.~A.; Papavassiliou,~G.~C. Hybrid Molecular Materials Based upon
  Organic $\pi$-Electron Donors and Inorganic Metal Complexes. Conducting Salts
  of Bis(ethylenediseleno)tetrathiafulvalene (BEST) with the Octahedral Anions
  Hexacyanoferrate(III) and Nitroprusside. \emph{Journal of Solid State
  Chemistry} \textbf{2002}, \emph{168}, 616--625\relax
\mciteBstWouldAddEndPuncttrue
\mciteSetBstMidEndSepPunct{\mcitedefaultmidpunct}
{\mcitedefaultendpunct}{\mcitedefaultseppunct}\relax
\EndOfBibitem
\bibitem[Kini \latin{et~al.}(1990)Kini, Geiser, Wang, Carlson, Williams, Kwok,
  Vandervoort, Thompson, Stupka, Jung, and Whangbo]{Kini1990}
Kini,~A.~M.; Geiser,~U.; Wang,~H.~H.; Carlson,~K.~D.; Williams,~J.~M.;
  Kwok,~W.~K.; Vandervoort,~K.~G.; Thompson,~J.~E.; Stupka,~D.~L.; Jung,~D.;
  Whangbo,~M.-H. {A New Ambient-Pressure Organic Superconductor,
  $\kappa$-(ET)$_2$Cu[N(CN)$_2$]Br, with the Highest Transition Temperature Yet
  Observed (Inductive Onset $T_c$ = $11.6$~K, Resistive Onset = $12.5$~K)}.
  \emph{Inorganic Chemistry} \textbf{1990}, \emph{29}, 2555--2557\relax
\mciteBstWouldAddEndPuncttrue
\mciteSetBstMidEndSepPunct{\mcitedefaultmidpunct}
{\mcitedefaultendpunct}{\mcitedefaultseppunct}\relax
\EndOfBibitem
\bibitem[Yagubskii \latin{et~al.}(1984)Yagubskii, Shchegolev, Laukhin,
  Kononovich, Kartsovnik, Zvarykina, and Buravov]{Yagubskii1984}
Yagubskii,~E.~B.; Shchegolev,~I.~F.; Laukhin,~V.; Kononovich,~P.;
  Kartsovnik,~M.; Zvarykina,~A.; Buravov,~L. {Normal-Pressure superconductivity
  in an organic metal (BEDT-TTF)$_2$I$_3$ [bis (ethylene dithiolo)
  tetrathiofulvalene triiodide]}. \emph{JETP Lett.} \textbf{1984}, \emph{39},
  12--16\relax
\mciteBstWouldAddEndPuncttrue
\mciteSetBstMidEndSepPunct{\mcitedefaultmidpunct}
{\mcitedefaultendpunct}{\mcitedefaultseppunct}\relax
\EndOfBibitem
\bibitem[Geiser \latin{et~al.}(1991)Geiser, Schults, Wang, Watkins, Stupka,
  Williams, Schirber, Overmyer, Jung, Novoa, and Whangbo]{Geiser1991}
Geiser,~U.; Schults,~A.~J.; Wang,~H.~H.; Watkins,~D.~M.; Stupka,~D.~L.;
  Williams,~J.~M.; Schirber,~J.; Overmyer,~D.; Jung,~D.; Novoa,~J.;
  Whangbo,~M.-H. {Strain index, lattice softness and superconductivity of
  organic donor-molecule salts}. \emph{Physica C: Superconductivity}
  \textbf{1991}, \emph{174}, 475--486\relax
\mciteBstWouldAddEndPuncttrue
\mciteSetBstMidEndSepPunct{\mcitedefaultmidpunct}
{\mcitedefaultendpunct}{\mcitedefaultseppunct}\relax
\EndOfBibitem
\bibitem[Yamochi \latin{et~al.}(1993)Yamochi, Komatsu, Matsukawa, Saito, Mori,
  Kusunoki, and Sakaguchi]{Yamochi1993}
Yamochi,~H.; Komatsu,~T.; Matsukawa,~N.; Saito,~G.; Mori,~T.; Kusunoki,~M.;
  Sakaguchi,~K. {Structural aspects of the ambient-pressure BEDT-TTF
  superconductors}. \emph{Journal of the American Chemical Society}
  \textbf{1993}, \emph{115}, 11319--11327\relax
\mciteBstWouldAddEndPuncttrue
\mciteSetBstMidEndSepPunct{\mcitedefaultmidpunct}
{\mcitedefaultendpunct}{\mcitedefaultseppunct}\relax
\EndOfBibitem
\bibitem[Elstner \latin{et~al.}(1993)Elstner, Singh, and Young]{Elstner1993}
Elstner,~N.; Singh,~R. R.~P.; Young,~A.~P. {Finite temperature properties of
  the spin-1/2 Heisenberg antiferromagnet on the triangular lattice}.
  \emph{Physical Review Letters} \textbf{1993}, \emph{71}, 1629--1632\relax
\mciteBstWouldAddEndPuncttrue
\mciteSetBstMidEndSepPunct{\mcitedefaultmidpunct}
{\mcitedefaultendpunct}{\mcitedefaultseppunct}\relax
\EndOfBibitem
\bibitem[Kanoda(2006)]{Kanoda2006}
Kanoda,~K. {Metal^^e2^^80^^93Insulator Transition in $\kappa$-(ET)$_2X$ and
  (DCNQI)$_2$ M: Two Contrasting Manifestation of Electron Correlation}.
  \emph{Journal of the Physical Society of Japan} \textbf{2006}, \emph{75},
  051007\relax
\mciteBstWouldAddEndPuncttrue
\mciteSetBstMidEndSepPunct{\mcitedefaultmidpunct}
{\mcitedefaultendpunct}{\mcitedefaultseppunct}\relax
\EndOfBibitem
\bibitem[Lee and Lee(2005)Lee, and Lee]{Lee2005}
Lee,~S.-S.; Lee,~P.~A. {U(1) Gauge Theory of the Hubbard Model: Spin Liquid
  States and Possible Application to $\kappa$-(BEDT-TTF)$_2$Cu$_2$(CN)$_3$}.
  \emph{Physical Review Letters} \textbf{2005}, \emph{95}, 036403\relax
\mciteBstWouldAddEndPuncttrue
\mciteSetBstMidEndSepPunct{\mcitedefaultmidpunct}
{\mcitedefaultendpunct}{\mcitedefaultseppunct}\relax
\EndOfBibitem
\bibitem[Motrunich(2005)]{Motrunich2005}
Motrunich,~O.~I. {Variational study of triangular lattice spin-1/2 model with
  ring exchanges and spin liquid state in $\kappa$-(ET)$_2$Cu$_2$(CN)$_3$}.
  \emph{Physical Review B - Condensed Matter and Materials Physics}
  \textbf{2005}, \emph{72}, 1--7\relax
\mciteBstWouldAddEndPuncttrue
\mciteSetBstMidEndSepPunct{\mcitedefaultmidpunct}
{\mcitedefaultendpunct}{\mcitedefaultseppunct}\relax
\EndOfBibitem
\bibitem[Watanabe \latin{et~al.}(2012)Watanabe, Yamashita, Tonegawa, Oshima,
  Yamamoto, Kato, Sheikin, Behnia, Terashima, Uji, Shibauchi, and
  Matsuda]{Watanabe2012}
Watanabe,~D.; Yamashita,~M.; Tonegawa,~S.; Oshima,~Y.; Yamamoto,~H.~M.;
  Kato,~R.; Sheikin,~I.; Behnia,~K.; Terashima,~T.; Uji,~S.; Shibauchi,~T.;
  Matsuda,~Y. {Novel pauli-paramagnetic quantum phase in a Mott insulator}.
  \emph{Nature Communications} \textbf{2012}, \emph{3}, 1--6\relax
\mciteBstWouldAddEndPuncttrue
\mciteSetBstMidEndSepPunct{\mcitedefaultmidpunct}
{\mcitedefaultendpunct}{\mcitedefaultseppunct}\relax
\EndOfBibitem
\bibitem[Furukawa \latin{et~al.}(2015)Furukawa, Miyagawa, Itou, Ito, Taniguchi,
  Saito, Iguchi, Sasaki, and Kanoda]{Furukawa2015}
Furukawa,~T.; Miyagawa,~K.; Itou,~T.; Ito,~M.; Taniguchi,~H.; Saito,~M.;
  Iguchi,~S.; Sasaki,~T.; Kanoda,~K. {Quantum Spin Liquid Emerging from
  Antiferromagnetic Order by Introducing Disorder}. \emph{Physical Review
  Letters} \textbf{2015}, \emph{115}, 077001\relax
\mciteBstWouldAddEndPuncttrue
\mciteSetBstMidEndSepPunct{\mcitedefaultmidpunct}
{\mcitedefaultendpunct}{\mcitedefaultseppunct}\relax
\EndOfBibitem
\bibitem[Yasaka \latin{et~al.}(2024)Yasaka, Yoshida, Tanaka, Nakamura, Kishida,
  Kitagawa, and Maesato]{Yasaka2024}
Yasaka,~S.; Yoshida,~Y.; Tanaka,~Y.; Nakamura,~Y.; Kishida,~H.; Kitagawa,~H.;
  Maesato,~M. {Electron Localization Induced by Disordered Anions in an Organic
  Conductor}. \emph{Inorganic Chemistry} \textbf{2024}, \emph{63},
  4196--4203\relax
\mciteBstWouldAddEndPuncttrue
\mciteSetBstMidEndSepPunct{\mcitedefaultmidpunct}
{\mcitedefaultendpunct}{\mcitedefaultseppunct}\relax
\EndOfBibitem
\bibitem[Florens and Georges(2002)Florens, and Georges]{Florens2002}
Florens,~S.; Georges,~A. Quantum impurity solvers using a slave rotor
  representation. \emph{Phys. Rev. B} \textbf{2002}, \emph{66}, 165111\relax
\mciteBstWouldAddEndPuncttrue
\mciteSetBstMidEndSepPunct{\mcitedefaultmidpunct}
{\mcitedefaultendpunct}{\mcitedefaultseppunct}\relax
\EndOfBibitem
\bibitem[Dasari \latin{et~al.}(2017)Dasari, Vidhyadhiraja, Jarrell, and
  McKenzie]{Dasari2017}
Dasari,~N.; Vidhyadhiraja,~N.~S.; Jarrell,~M.; McKenzie,~R.~H. {Quantum
  critical local spin dynamics near the Mott metal-insulator transition in
  infinite dimensions}. \emph{Physical Review B} \textbf{2017}, \emph{95},
  1--8\relax
\mciteBstWouldAddEndPuncttrue
\mciteSetBstMidEndSepPunct{\mcitedefaultmidpunct}
{\mcitedefaultendpunct}{\mcitedefaultseppunct}\relax
\EndOfBibitem
\bibitem[Isono \latin{et~al.}(2016)Isono, Terashima, Miyagawa, Kanoda, and
  Uji]{Isono2016}
Isono,~T.; Terashima,~T.; Miyagawa,~K.; Kanoda,~K.; Uji,~S. {Quantum
  criticality in an organic spin-liquid insulator
  $\kappa$-(BEDT-TTF)$_2$Cu$_2$(CN)$_3$}. \emph{Nature Communications}
  \textbf{2016}, \emph{7}, 13494\relax
\mciteBstWouldAddEndPuncttrue
\mciteSetBstMidEndSepPunct{\mcitedefaultmidpunct}
{\mcitedefaultendpunct}{\mcitedefaultseppunct}\relax
\EndOfBibitem
\bibitem[Riedl \latin{et~al.}(2019)Riedl, Valent{\'{i}}, and Winter]{Riedl2019}
Riedl,~K.; Valent{\'{i}},~R.; Winter,~S.~M. {Critical spin liquid versus
  valence-bond glass in a triangular-lattice organic antiferromagnet}.
  \emph{Nature Communications} \textbf{2019}, \emph{10}, 2561\relax
\mciteBstWouldAddEndPuncttrue
\mciteSetBstMidEndSepPunct{\mcitedefaultmidpunct}
{\mcitedefaultendpunct}{\mcitedefaultseppunct}\relax
\EndOfBibitem
\bibitem[Mori and Katsuhara(2002)Mori, and Katsuhara]{Mori2002}
Mori,~T.; Katsuhara,~M. {Estimation of $\pi$d-Interactions in Organic
  Conductors Including Magnetic Anions}. \emph{Journal of the Physical Society
  of Japan} \textbf{2002}, \emph{71}, 826--844\relax
\mciteBstWouldAddEndPuncttrue
\mciteSetBstMidEndSepPunct{\mcitedefaultmidpunct}
{\mcitedefaultendpunct}{\mcitedefaultseppunct}\relax
\EndOfBibitem
\bibitem[Ito \latin{et~al.}(2022)Ito, Kobayashi, Sari, Watanabe, Saito,
  Kawamoto, Tsunakawa, Satoh, and Taniguchi]{Ito2022}
Ito,~A.; Kobayashi,~T.; Sari,~D.~P.; Watanabe,~I.; Saito,~Y.; Kawamoto,~A.;
  Tsunakawa,~H.; Satoh,~K.; Taniguchi,~H. {Antiferromagnetic ordering of
  organic Mott insulator $\lambda$-(BEDSe-TTF)$_2$GaCl$_4$}. \emph{Physical
  Review B} \textbf{2022}, \emph{106}, 045114\relax
\mciteBstWouldAddEndPuncttrue
\mciteSetBstMidEndSepPunct{\mcitedefaultmidpunct}
{\mcitedefaultendpunct}{\mcitedefaultseppunct}\relax
\EndOfBibitem
\bibitem[Foury-Leylekian \latin{et~al.}(2018)Foury-Leylekian, Ilakovac,
  Bal{\'{e}}dent, Fertey, Arakcheeva, Milat, Petermann, Guillier, Miyagawa,
  Kanoda, Alemany, Canadell, Tomic, Pouget, Foury-Leylekian, Ilakovac,
  Bal{\'{e}}dent, Fertey, Arakcheeva, Milat, Petermann, Guillier, Miyagawa,
  Kanoda, Alemany, Canadell, Tomic, and Pouget]{Foury-Leylekian2018}
Foury-Leylekian,~P. \latin{et~al.}  {(BEDT-TTF)$_2$Cu$_2$(CN)$_3$ Spin Liquid:
  Beyond the Average Structure}. \emph{Crystals} \textbf{2018}, \emph{8},
  158\relax
\mciteBstWouldAddEndPuncttrue
\mciteSetBstMidEndSepPunct{\mcitedefaultmidpunct}
{\mcitedefaultendpunct}{\mcitedefaultseppunct}\relax
\EndOfBibitem
\end{mcitethebibliography}
\end{document}